\documentclass[aps,twocolumn,showpacs,preprintnumbers,amsmath,amssymb,nofootinbib,superscriptaddress,showkeys]{revtex4-1}
\usepackage{graphicx,graphics,epsfig}
\usepackage{amsmath,amssymb,slashed,dcolumn,amsfonts,slashed,tabulary}
\usepackage{booktabs}
\usepackage{comment}
\usepackage{multirow}

\newcommand{\beq}{\begin{equation}}
\newcommand{\eeq}{\end{equation}}
\newcommand{\beqa}{\begin{eqnarray}}
\newcommand{\eeqa}{\end{eqnarray}}
\newcommand{\bal}{\begin{align}}

\begin{document}

\title{Self-consistent statistical error analysis of $\pi\pi$
  scattering}

\author{R. Navarro P\'erez}\email{rnavarrop@ugr.es}
\author{E. Ruiz Arriola}\email{earriola@ugr.es}
\affiliation{Departamento de
    F\'{\i}sica At\'omica, Molecular y Nuclear and Instituto Carlos I
    de F{\'\i}sica Te\'orica y Computacional \\ Universidad de
    Granada, E-18071 Granada, Spain.}  \author{J. Ruiz de
    Elvira}\email{elvira@hiskp.uni-bonn.de}
  \affiliation{Helmholtz-Institut f\"ur Strahlen- und Kernphysik
    (Theorie) and Bethe Center for Theoretical Physics,
    \\ Universit\"at Bonn, Germany }

\date{\today}

\begin{abstract} 
\rule{0ex}{3ex} We analyze the conditions under which a statistical
error analysis can be carried out in the case of $\pi\pi$ scattering,
namely the normality of residuals in the conventional $\chi^2$-fit
method. Here we check that the current and benchmarking analyses only present very small
violations of the normality requirements. 
In particular, we show how it is possible to amend
slightly the selection of the experimental data, and improve the normality of residuals.
 As an example, we discuss
the $0^{++}$ channel and the implications for the $f_0(500)$ and
$f_0(980)$ resonances, obtaining that the new selection of data
provides very similar and compatible results.  In addition, the
effect on the $f_0(500)$ and $f_0(980)$ resonance pole parameters is
almost negligible, which reinforces the central results and the
uncertainty analysis performed in these benchmarking determinations.

\end{abstract}


\maketitle

\section{Introduction}

Pion-pion scattering is the simplest hadronic reaction in
QCD where most theoretical control based on analyticity, crossing,
chiral symmetry and Regge behavior has been
achieved (for a pedagogical introduction to this topic see for example~\cite{Martin:1976mb}). 
The most impressive consequence of this development has been the determination of S-wave scattering lengths
with an order of magnitude more precise than the experimental
determination~\cite{Colangelo:2000jc,Ananthanarayan:2000ht,Colangelo:2001df,Caprini:2003ta,Pelaez:2004vs,Kaminski:2006yv,Kaminski:2006qe,GarciaMartin:2011cn},
an exceptional case in strong interactions where most often just the
opposite situation is encountered, and experimental precision takes
over theoretical predictive power.

Furthermore, the mass and the width of the lowest resonance in the
$0^{++}$ channel, known as $f_0(500)$, and lodged in the
PDG~\cite{Agashe:2014kda} to stay, have been pinned down unambiguously
and accurately~\cite{Caprini:2005zr,GarciaMartin:2011jx} after a long
history full of debates. The scalar meson was already proposed by
Johnson and Teller in 1955~\cite{Johnson:1955zz} as the mid-range
mediator of the nuclear force necessary for nuclear binding of atomic
nuclei (a historic account on the appearance and disappearance of the
$\sigma-$meson in particle physics may be traced from the
current~\cite{Agashe:2014kda} and previous editions of the PDG
booklet). Final values have been reported which agree within
uncertainties. This high precision has been a novel and key convincing
culminating element in favor of a precise determination of the $f_0(500)$ pole parameters. 
Despite of the fact that systematic uncertainties often dominate $\pi\pi$-scattering analyses,
the application of statistical techniques such as the
well-known least squares $\chi^2$ method has been massive and a
crucial element to quote uncertainties estimates.

We remind that the least squares method rests on the major assumption
that discrepancies between the most likely theory and the experiment
are statistical fluctuations and more specifically independent random
normal variables. If there are good reasons to suspect this
assumption, one cannot assume the standard approach for error
propagation. Thus, in order to be able to use $\chi^2$ as a means to
obtain uncertainties, this expected normality of residuals must and
can be checked {\it a posteriori}. Of course, this self-consistency of
the fitting procedure cannot be decided with absolute certainty, due
to the finite amount of the fitted data. Thus, the key question is to
decide whether a given finite set of residuals follow, within expected
statistical fluctuations, a normal distribution pattern or not. The
verification of this important condition is termed normality test of
residuals and, though elementary, it has been completely disregarded
in all previous experimental and theoretical studies in
$\pi\pi$-scattering.  This is in contrast with the painstaking efforts
to reduce errors taken almost elsewhere in these works.

We remind that the theoretical modeling of experimental data requires
within the classical statistical setup an assumption both for the
signal (the fitting curve) and the noise (the statistical
fluctuation). However, for a self-consistent treatment both
assumptions may always be checked a posteriori and, if corroborated,
error propagation may then be undertaken according to the verified
distribution of statistical (not necessarily normal) fluctuations. A
particular and important example of this is provided by the standard
and often recommended practice of re-scaling (by a Birge
factor~\cite{birge1932calculation,kacker2008classical}) of a too large
$\chi^2$ which corresponds to artificially and globally enlarge the
errors when incompatible data are detected among different sets of
measurements~\cite{Beringer:1900zz,behnke2013data}).  This procedure
is only justified when we can confidently state that the residuals are
distributed as scaled gaussians. In such a positive case, we can still
propagate errors according to the {\it re-scaled} gaussian
distribution.  If, however, this is not the case, enlarging the errors
in the experiment does not entitle to propagate errors according to
some distribution.

In the present paper we want to scrutinize the consistency between
$\pi\pi$ scattering data used as the experimental input information
and the theoretical framework used to analyze it. This will be done
according to the standard rules of the statistical analysis, namely by
inquiring whether or not the difference between the measured
experimental data and the theory being tested can be assumed to be a
fluctuation which would likely decrease with further and more accurate
measurements~\footnote{We note that much used routines such as
  MINUIT~\cite{james2004minuit} do not implement this necessary and
  simple test and actually very few textbooks currently used by
  particle physicists mention this important
  requirement~\cite{eadie2006statistical} (one exception is
  Ref.~\cite{evans2004probability}).}.

As it was previously analyzed in \cite{Pelaez:2004vs}, there is some
tension among the available experimental $\pi\pi$ data.  This
suggests, as it is usually the case, to make a judicious decision on
which data should be selected as plausibly consistent and,
consequently, which data should be discarded as inconsistent. This
decision embodies a certain probability of making an erroneous choice,
a number which can be estimated quantitatively on the basis of the
assumed model. One naturally expects that the data selection improves
the statistical behavior of the residuals, in which case it seems
unlikely that the theoretical model contains significant systematic
errors. This implies in turn that the uncertainties in the model
parameters faithfully reflect the uncertainties of the selected and
mutually compatible data.

Our aim here is to restrict the discussion to the simple $0^{++}$
channel as an example where the necessary statistical tools can be
presented, and to use the Madrid-Krakow $\pi\pi$
analysis~\cite{Pelaez:2004vs,Kaminski:2006yv,Kaminski:2006qe,GarciaMartin:2011cn}
as illustration, taking advantage that one of us (J.R.E)
was one of the authors of this collaboration. 
This procedure can be used as an initial guess of a
more comprehensive analysis based on solving Roy and Roy-like
equations supplemented with forward dispersion relations and sum
rules.  As we will show, our results based in our restricted analysis
will indicate that, after the statistical self consistency is over
imposed, there are no dramatic changes to be expected,
providing a support of the results of~\cite{GarciaMartin:2011cn,GarciaMartin:2011jx}.
Due to the lack of popularity, our presentation is intentionally pedagogical

The paper is organized as follows. In Sect.~\ref{sec:old-UFD} we
review the latest $\pi\pi$ analysis performed by the Madrid-Krakow group~\cite{GarciaMartin:2011cn}. 
Next, in Sect.~\ref{sec:norm-test}, we review the notion of normality test
as well as some other useful tests. In Sect.~\ref{sec:testing} we apply
these normality tests to the S0 wave of the $\pi\pi$ analysis in \cite{GarciaMartin:2011cn}.  
In Sect.~\ref{sec:new-UFD} we show how suitably discarding data results
regarding normality tests improve. In Sect.~\ref{sec:self-consistent}
we make a first attempt on making a $3\sigma$ self-consistent fit of
mutually compatible data. Finally, in Sect.~\ref{sec:concl} we draw
our main conclusions and provide an outlook for future work.

\section{Review on the Unconstrained Fit to Data (UFD)}
\label{sec:old-UFD}

In a series of
works~\cite{Pelaez:2004vs,Kaminski:2006yv,Kaminski:2006qe,GarciaMartin:2011cn},
the Madrid-Krakow group has used a dispersive approach to built a
$\pi\pi$-scattering amplitude which incorporates analyticity,
unitarity and crossing symmetry. As a starting point, a set of simple
expressions parametrizing the available experimental data in each
partial wave amplitude were fitted to data independently. This is
known as Unconstrained Fit to Data (UFD).  These parametrizations were
then checked against dispersion relations. When the answer was in the
affirmative, these parametrizations were used as a starting point for
a Constrained Fit to Data (CFD), in which dispersion relations were
imposed as an additional constraint. Therefore, the set of parameters
obtained describe $\pi\pi$-scattering consistently with dispersion
relations. Finally, these parametrizations were then used as input for
the dispersive Roy and GKPY equations to perform an analytic
extrapolation to the complex plane, yielding precise and model
independent determinations of the lightest resonance poles appearing
in $\pi\pi$-scattering~\cite{GarciaMartin:2011jx}.

However, one potential objection to these works, 
already noted by the authors,  concerns the uncertainty treatment.  
In particular in~\cite{GarciaMartin:2011cn}, uncertainties were calculated
following two approaches.  On the one hand, the effect of varying each
parameter independently in the parametrizations from $p_i$ to $p_i\pm
\Delta p_i$ was added in quadrature, disregarding possible
correlations among them. The errors were further symmetrized to the
largest variation $\Delta p_i$ conservatively covering the confidence
interval. Alternatively, errors were also estimated using a Monte
Carlo Gaussian sampling~\cite{MonteCarlo} (see
e.g. \cite{Nieves:1999zb} for early implementations of these ideas in
$\pi\pi$ scattering) of all UFD parameters (within 6 standard
deviations). The uncertainties defined by excluding $16\%$ of the
upper and lower tails of the distribution of a total of $10^5$ events
turned out be slightly asymmetric. While the essence of the Monte
Carlo method is to keep correlations, final results were not very
different from the most naive approach. This was partly due to the
smallness of errors as well as the large number of parameters.

On the other hand, the way the experimental data are represented via
standard least squares $\chi^2$ fitting procedures where one minimizes:  
\begin{eqnarray}
\chi^2 ({\bf p}) = \sum_{i=1}^N \left( \frac{{\cal O}_i^{\rm exp}- {\cal O}_i^{\rm th} ( {\bf p
})}{\Delta {\cal O}_i^{\rm exp}} \right)^2,
\end{eqnarray}
where $O_i^{\rm exp}$ are experimental observations  with estimated
errors $\Delta O_i^{\rm exp}$,  ${\cal O}_i({\bf p})$ is the theoretical model  with $P$ parameters ${\bf p}=(p_1, \dots, p_P)$ and $N$ the number of data points. At the minimum,
\begin{eqnarray}
\chi^2_{\rm min} \equiv \min_{\bf p} \chi^2({\bf p}) = \chi^2 ({\bf p}_0) \, , 
\end{eqnarray}
the most likely theory parameters are ${\bf p}_0$, and the most likely
theoretical prediction is ${\cal O}_i^{\rm th} \equiv O_i({\bf p}_0).$
The residuals at ${\bf p}_0$ are defined by:
\begin{eqnarray}\label{residual-definition}
R_i  = \frac{{\cal O}_i^{\rm exp}-{\cal O}_i({\bf p}_0)}{\Delta {\cal O}_i^{\rm exp}}.
\end{eqnarray}
and one requires the residuals $R_i$ to be normally
distributed~\cite{Perez:2014yla,Perez:2014kpa}~\footnote{These works
  refer to Nucleon-Nucleon scattering where a set of about 8000
  neutron-proton and proton-proton scattering data below pion
  production threshold and published from 1950 till 2013 have been
  comprehensively analyzed along similar lines where many more details
  can be found. In order to spare the effort of the reader unfamiliar
  with the NN problem we try to make the presentation here
  sufficiently self-contained.}. When normality is fulfilled, 
the $\chi^2$ test requires (for large $N$) that within $1\sigma$
$\chi^2_{\rm min}/\nu= 1 \pm \sqrt{2/\nu}$, with $\nu=N-P$ the numbers of degrees of freedom (dof). 
Therefore, this requirement precludes {\it both} loo large {\it and} and too small 
$\chi^2/{\rm dof}$. 

In this section we describe the fitting procedure in detail and
postpone the normality analysis for the next Section.

\subsection{S0 wave parametrization}

Let us first review briefly the different parametrizations used
in~\cite{GarciaMartin:2011cn} for the S0 wave.  The partial-wave is
written as follows,
\begin{eqnarray}
t^{(0)}_0(s)=\frac{\sqrt{s}}{2 k}\hat f^{(0)}_0(s), \quad
\hat f^{(0)}_0(s)=\frac{\eta_0^{(0)}(s)e^{2i\delta_0^{(0)}(s)}-1}{2i},
\nonumber 
\end{eqnarray}
where $\delta_0^{(0)}(s)$ and $\eta_0^{(0)}(s)$ are the phase shift
and inelasticity of the partial wave, and $k=\sqrt{s/4-m_\pi^2}$ is the center of mass
momentum.

At low energies, where the elastic approximation is valid,
a model-independent parametrization ensuring elastic unitarity is used, 
\begin{eqnarray}
t_0^{(0)}=\frac{\sqrt{s}}{2 k}\frac{1}{\cot\delta_0^{(0)}(s)-i}.
  \label{flI}
\end{eqnarray}
In addition, in order to ensure maximal analyticity in the complex plane,
$\cot\delta_0^{(0)}(s)$ is expanded in powers of the conformal variable $w(s)$,
\begin{eqnarray}
\label{eq:S0lowparam}
&&  \cot\delta_0^{(0)}(s)=
\frac{s^{1/2}}{2k}\frac{M_\pi^2}{s-\frac{1}{2}z_0^2} \times \nonumber \\
&&\left(\frac{z_0^2}{M_\pi\sqrt{s}}+B_0+B_1w(s)+B_2w(s)^2+B_3w(s)^3\right),
\nonumber\\
&&w(s)=\frac{\sqrt{s}-\sqrt{s_0-s}}{\sqrt{s}+\sqrt{s_0-s}},
\qquad s_0=4M_K^2.
\end{eqnarray}
In the intermediate energy inelastic region, a purely polynomial
expansion is used for the phase shift.  In addition, continuity, and a
continuous derivative are imposed at the matching point, chosen at
$s_M^{1/2}=850$ MeV,
\begin{eqnarray}\label{eqpol}
&\delta_0^{(0)}(s)=\left\{ 
\begin{array}{l}
d_0\left(1-\frac{\vert k_2 \vert}{k_M}\right)^2+\delta_M \, \frac{\vert k_2 \vert}{k_M}\left(2-\frac{\vert k_2 \vert}{k_M}\right) \\  
\\
 +\vert k_2 \vert (k_M-\vert k_2 \vert)
\left(8\delta_M'+c \frac{(k_M-\vert k_2 \vert)}{M_K^3}\right) \, , \\
\\ 
\qquad  (0.85 \,{\rm GeV})^2<s<4 M_K^2, \\
 \\
d_0+B\frac{k_2^2}{M_K^2}+C\frac{k_2^4}{M_K^4}+D\,\theta(s-4M_\eta^2)\frac{k_3^2}{M_\eta^2} \, ,  \\  
\\
\qquad 4 M_K^2<s<(1.42 \mathrm{GeV})^2,
\end{array}\right.
\end{eqnarray} 
where $k_2=\sqrt{s/4-M_K^2}$. Note that we have defined
$k_M=\vert k_2(s_M), \vert$$\delta_M=\delta(s_M)$ and $\delta_M'=d\delta(s_M)/ds$, which are
obtained from Eq.~\eqref{eq:S0lowparam}.

Finally, an elastic S0 wave, $\eta_0^{(0)}=1$, up to the two-kaon
threshold is assumed, whereas above that energy, it is parametrized
as:
\begin{eqnarray}
\eta_0^{(0)}(s)&=&\exp\bigg[\frac{-k_2(s)}{s^{1/2}}\left(\tilde\epsilon_1+
\tilde\epsilon_2\,
\frac{k_2}{s^{1/2}} +\tilde\epsilon_3\,\frac{k_2^2}{s}\right)^2 \nonumber \\ 
&&-\tilde\epsilon_4 \theta(s-4M_\eta^2)\frac{k_3(s)}{s^{1/2}}\bigg].
\label{eq:inelasticityS0}
\end{eqnarray}

As we will review in the following subsections, 
the number of data points used in~\cite{GarciaMartin:2011cn} for each parametrization is $N=30,61,28$
respectively. In addition, the number of parameters of each of them is $P=4,5,4$. 
Thus, it means that to $1\sigma$ we expect at
the minimum $\chi^2/\nu = 1 \pm \sqrt{2/\nu}= 1 \pm 0.28, 1 \pm 0.19,
1 \pm 0.28$ with $\nu = N-P$ respectively. 
Therefore, the total value is expected to be $\chi^2/\nu = 1 \pm 0.14$.
In the following, we will review the different sets of data
considered in~\cite{GarciaMartin:2011cn} to fit each of these parametrizations.

\subsection{Low energy region: conformal fit}

For this fit, it was considered data from the pion threshold up to
$M_K$ on $K_{\ell l4}$ decays~\cite{Rosselet:1976pu}, including the latest
data from NA48/2~\cite{ULTIMONA48}, 
and the selection performed in~\cite{Pelaez:2004vs} of all the existing and conflicting $\pi\pi$ scattering
data~\cite{pipidata,Grayer,Hyams75} between 800 MeV and the matching
point at 850 MeV (PY05). This selection was obtained from an average of the
different experimental sets passing a consistency test with
Forward Dispersion Relations (FDR), 
where the uncertainties were chosen so that they
covered the difference between the initial data sets.
A detailed description of the data selection considered in this average  
can be found in~\cite{Pelaez:2004vs} or in the Appendix of Ref.~\cite{Yndurain:2007qm}. 

\begin{figure}
  \begin{center}
  \hspace{0cm} \includegraphics[scale=0.2]{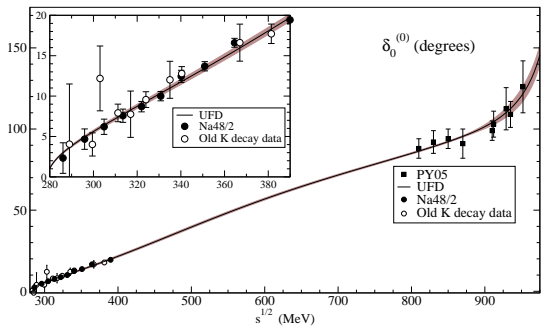}
  \end{center}
  \caption{UFD for the $\pi\pi$ S0 wave. The dark band covers the uncertainties,
    versus the fitted data, namely, the ``old'' phase shift data from
    $K_{\ell4}$ decays~\cite{Rosselet:1976pu}, the final NA48/2
    results~\cite{ULTIMONA48} and the data average
    performed in PY05~\cite{Pelaez:2004vs}.
    \label{fig:phaselowenergies}}
\end{figure}
In Table~\ref{tab:Conformal-residuals}, we show the experimental data
considered, together with the fitted values and their corresponding
residuals. In addition, in Fig.~\ref{fig:phaselowenergies}, we show the low energy
conformal fit in detail, and in Table~\ref{tab:S0parameters}, we show
the UFD parameters obtained from the fit.

\subsection{Intermediate energy region: polynomial fit}

\begin{figure}[h]
  \begin{center}
   \hspace{0cm}\includegraphics[scale=0.25]{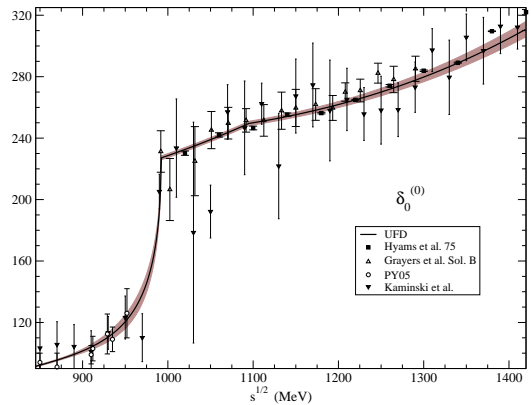}
  \end{center}
  \caption{S0 UFD, where the dark band covers the uncertainties,
    versus the fitted data, namely, the data average performed in
    PY05~\cite{Pelaez:2004vs}, the re-analysis of CERN-Munich
    experiments~\cite{Kaminski}, and data from~\cite{Grayer,Hyams75}.
    \label{fig:phaseintermidiate}}
\end{figure}

This parametrization describes the S0 wave phase shift in the energy region 
between 850 and 1420 MeV.  Below the $K\bar K$ threshold, it is fitted to the PY05
average of the different experimental solutions compatible with
FDR~\cite{Pelaez:2004vs}, and to the re-analysis of the CERN-Munich
experiments~\cite{pipidata} performed by Kaminski et
al.~\cite{Kaminski}, which includes a huge estimation of the possible
statistical uncertainties.  Above the $K\bar K$ threshold, only were
considered data sets with inelasticities compatible with
$\pi\pi\rightarrow K\bar K$ scattering results in the region
$4M_K^2\leq s \leq (1.25\,{\rm GeV})^2$~\cite{Wetzel:1976gw}, namely,
the solution $(-\,-\,-)$ of Hyams et al.~\cite{Hyams75}, data of
Grayer et al.~\cite{Grayer} and data of Kaminski et
al.\cite{Kaminski}.  Unfortunately, data coming from~\cite{Hyams75}
only provided a statistical uncertainty estimation, so $5^\circ$ were
added in~\cite{GarciaMartin:2011cn} as systematic error.  In Fig.~\ref{fig:phaseintermidiate}, we
show the resulting polynomial fit and in Table~\ref{tab:S0parameters},
the UFD parameters obtained from the fit.
In addition, as we did for the previous fit, in
Table~\ref{tab:polynomial-residuals}, we present the experimental data
considered, together with the fitted values and their corresponding
residuals.

\subsection{S0 inelasticity fit}

For the inelasticity, only data consistent with the dip solution were
considered, namely the 1973 data of Hyams et al.~\cite{pipidata},
Protopopescu et al.~\cite{pipidata} and Kaminski et
al.{\cite{Kaminski}.  However, the data from Kaminski et al. were not
  included in the fit.  As explained in~\cite{GarciaMartin:2011cn},
  the reason behind this is that the main source of uncertainty was
  systematic, so the large number of points of Kaminski et
  al. together with the huge statistical errors would lead to a fit
  outcome with much smaller errors than the original systematic
  uncertainties.  With only the other two sets, which are
  incompatible, a fit with a large $\chi^2/d.o.f.$ was obtained, and
  by rescaling its uncertainties in the inelasticity parameters, it
  was possible to mimic the dominant systematic uncertainties much
  better. Of course, the results were still in very good agreement
  with Kaminski et al.  In Table~\ref{tab:S0parameters} of the
  Appendix, we provide the values for the $\tilde\epsilon_i$
  parameters, and in Fig.~\ref{fig:S0inel} we show the results of the
  unconstrained fit to the S0 wave inelasticity data up to 1420 MeV.
\begin{figure}[h]
  \begin{center}
  \hspace{0cm} \includegraphics[scale=0.25]{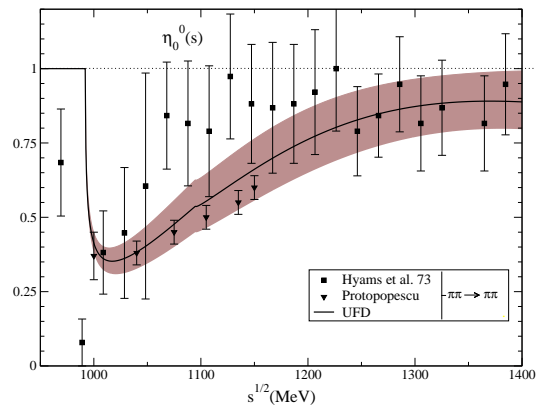}
  \end{center}
  \caption{ S0 inelasticity fit (UFD set) to the
    $\pi\pi\rightarrow\pi\pi$ scattering data of Hyams et al. (1973)
    and Protopopescu et al. As explained in the text, Kaminski et
    al. data were not fitted \protect{\cite{pipidata}} although the
    fit is compatible with them.  The dark band covers our
    uncertainties.
    \label{fig:S0inel}}
\end{figure}

\section{Normality test of the UFD fit}
\label{sec:norm-test}

\subsection{General remarks}

In order to understand the discussion below, we have to introduce some
well known concepts from the statistical theory of data analysis and
specifically on normality tests. While the ideas are easy to grasp,
these simple but extremely informative tests are too often overlooked
as to deserve to be reviewed in some detail. We avoid on purpose the
conventional statistical jargon which, after exchanging view with
other researchers, usually backs off theoretical physicists from going
in-depth through lengthy textbooks and use canned routines as a black
box. The reader familiar with normality tests may skip this section.

As it is well known, the least squares method implicitly assumes that
the residuals building the total $\chi^2$ are normally distributed. Of
course, for a finite number of data, we can never be certain about
this. We can however, estimate the probability by which we would err
when we decide that this is not the case. In case we have no
objection, we can plausibly assume that further measurements will
decrease the uncertainties in the fitting parameters; otherwise we
should expect incompatible determinations of fitting parameters. 

In our case we want to decide whether or not the set of $N$ residuals
$(R_1, \dots R_N)$ from the $\chi^2$-fit obey a standardized normal
distribution. The question is whether or not the deviations of this
empirical sample from the theoretical distribution can plausibly be
attributed to the fact that number of data points $N$ is finite. This can be
easily done by generating random numbers drawn from a normal
distribution $(\xi_1, \dots, \xi_N)$, $\xi_i \in N(0,1)$, and
comparing some quantity of interest $T(\xi_1, \dots, \xi_N)$ with the
actually observed empirical result $T_{\rm obs}=T(x_1, \dots,
x_N)$. Of course, we can repeat the process as many times $a=1,
\dots, M$ as we want and we will obtain a spread of results $T_{\rm
  th,a}=T(\xi_{1,a}, \dots, \xi_{N,a})$ yielding a distribution of
values for the variable $T$ assuming the variables $(\xi_{1,a}, \dots,
\xi_{N,a})$ are uncorrelated normal variables.

A very important issue is to agree on {\it what} quantity should be
used for this comparison. Let us consider, for instance, the moments
centered at the origin defined as:
\begin{eqnarray}
\mu_r(x_1, \dots, x_N)= \frac1N \sum_{i=1}^N x_i^r, 
\end{eqnarray}
Of course, for normally distributed variables $\mu_r(\xi_1, \dots,
\xi_N)$ becomes a random variable itself which follows a certain
distribution,
\begin{eqnarray}
P_N (\mu_r) = \langle \delta(\mu_r -\mu_r (\xi_1, \dots, \xi_N)) \rangle, 
\end{eqnarray}
where the expectation value is defined as: 
\begin{eqnarray}
\langle O\rangle_N  &=& \int_{-\infty}^{\infty} d\xi_1 \frac{e^{-\xi_1^2/2}}{\sqrt{2\pi}} 
\dots \int_{-\infty}^{\infty} d\xi_N \frac{e^{-\xi_N^2/2}}{\sqrt{2\pi}}  O(\xi_1, \dots, \xi_N). \nonumber \\ 
\end{eqnarray}
This function cannot be computed analytically, except for $r=1,2$ or
when $N \to \infty$ (featuring the central limit
theorem~\cite{evans2004probability}),
\begin{eqnarray}
  P_N (\mu_r) \to \frac{e^{-\frac12 \left[\frac{\mu_r - \langle \mu_r \rangle}{\Delta \mu_r}\right]^2 }}{\sqrt{2\pi} \Delta \mu_r},
\end{eqnarray}
where the mean $\langle \mu_r \rangle$ and the standard deviation
$(\Delta \mu_r)^2 = \langle \mu_r^2 \rangle - \langle \mu_r \rangle^2
$ can be expressed in terms of Euler's gamma
functions~\cite{evans2004probability}. The lowest values $r \le 8$ are
listed for a quick reference in Table~\ref{tab:moments}.  Thus, we
have that if $x_i \in N[0,1]$ then within $1\sigma$ ($68\%$)
confidence level and for large $N$,
\begin{eqnarray}
\frac1N \sum_{i=1}^N z_i^r  = \langle \mu_r \rangle \pm \Delta \mu_r.
\end{eqnarray}
When $N$ is not as large (say $\lesssim 30$) one can proceed by Monte
Carlo by sampling $N$ gaussian numbers and computing the distribution
of moments.

\begin{table}[h]
\caption{The normalized moments $\mu_r$ of the standardized gaussian
  distribution with their mean standard deviation $\Delta \mu_r$ for a
  sample of size $N$. For normal distributed data $x_i \in N(0,1)$
  we expect that to $1\sigma$ confidence level $\sum_{i=1}^N x_i^r/N=
  \mu_r \pm \Delta \mu_r$.}
\begin{ruledtabular}
  \begin{scriptsize}
\begin{tabular}{l|ccccccccc}
 $r$ & 0 & 1 & 2 & 3 & 4 & 5 & 6 & 7 & 8  \\
 $\mu_r$ & 1 & 0 & 1 & 0 & 3 & 0 & 15 & 0 & 105  \\
 $\Delta \mu_r$ & 0 & $\frac{1}{\sqrt{N}}$ & $\frac{\sqrt{2}}{\sqrt{N}}$ & $\frac{\sqrt{15}}{\sqrt{N}}$ & $\frac{4
   \sqrt{6}}{\sqrt{N}}$ & $\frac{3 \sqrt{105}}{\sqrt{N}}$ & $\frac{3 \sqrt{1130}}{\sqrt{N}}$ &
   $\frac{3 \sqrt{15015}}{\sqrt{N}}$ & $\frac{240 \sqrt{35}}{\sqrt{N}}$ 
\end{tabular}
  \end{scriptsize}
\end{ruledtabular}
 \label{tab:moments}
\end{table}

In the moments method we weight the tails more strongly as the power
$r$ increases. Therefore, rejecting or accepting normality on the
basis of each individual moment $\mu_r$ tests different features of
the distribution. This discussion shows that normality is to some
extent in the eyes of the beholder, and generally it is better to use
different tests.

As we have already mentioned, the modeling of data requires an
assumption both for the signal (the fitting curve) and the noise (the
statistical fluctuation). In this regard, there is an interesting
situation where the moments do not stem from a standardized normal
distribution, but after suitably shifting and re-scaling they do. In
such a case we can still propagate errors according to the shifted
and re-scaled distribution.

\subsection{Normality tests}\label{sect:normality-tests}

The normality test consists of an {\it a priori} criterium where one
decides if the set of data $(x_1 \dots, x_N)$ could possibly be
normal. To make this decision, one evaluates what is the probability
$p$ of making an erroneous decision of denying the normality of the
residuals. The $p$-value is obtained by locating an observable $T$,
known as {\it test statistic}, from the actual empirical data $(x_1,
\dots, x_N)$ in the theoretical distribution that this expected in
case the data do follow the normal distribution.  A small $p$-value
indicates clear deviations from the normal distribution, whereas a
large $p$-value indicates that no statistically significant
discrepancies were found.  When comparing $T_{\rm obs}$ to the
distribution of $T$ a \emph{significance level} $\alpha$ is
arbitrarily chosen, this determines a critical value $T_c$.  Common
choices for $\alpha$ are $0.01$ and $0.05$. We then compare the
observed $T$ with the critical theoretical value $T_c$. The definition
of $T$ in each test determines which inequality must be fulfilled
either from the left or the right. Therefore $T \ge T_c $~\footnote{We
  are assuming the most frequent case where large $T$ means large
  deviations from the theoretical distribution, as for instance in the
  Pearson and Kolmogorov-Smirnov tests (see below). The situation
  where $T \le T_c$ is the relevant inequality can also take place,
  for example in the Tail-sensitive test explained below.} the
assumption that the empirical data $ x_1, \dots , x_n $ were drawn
from the probability distribution $\rho(x)$ can be rejected with a
confidence level of $100(1-\alpha)\%$. If $T <  T_c$ there are no
statistically significant reasons to reject the assumption. This is
usually expressed in short saying that the finite sample $(x_1, \dots,
x_N)$ follows the normal distribution $N(0,1)$ with $100(1-\alpha)\%$
confidence level.

There are many tests available on the market, and we will take here
the most popular, which are the Pearson, Kolmogorov-Smirnov (KS) and
Tail-Sensitive (TS) tests. We will also use the moments method already
described above.

\subsubsection{Pearson test}

The Pearson test is one of the first tools used to test the goodness of fit
by comparing the histograms of the empirical data (the residuals of a
fit in this case) and of the normal distribution. For this purpose a
binning must be specified and the data is binned accordingly and the
test statistic is defined as
\begin{equation}
 T_{\rm Pearson} = \sum_{i=1}^{Nb} \frac{\left(n_i^{\rm res} - n_i^{\rm normal}\right)^2}{n_i^{\rm normal}},
\end{equation}
where $N_b$ is the number of bins, $n_i^{\rm fit}$ is the number of
residuals on each bin and $n_i^{\rm normal}$ the expected number of data
from a normal distribution on the same bin. In this case a large value
of $T_{\rm Pearson}$ indicates discrepancies between the empirical and
normal distribution and therefore the null hypothesis is rejected if
$T_{\rm obs} > T_c$.

One particular disadvantage of the Pearson test is the dependence of
$T_{\rm obs}$ on the number and size of the bins, which are set
arbitrarily.  Different choices can be made about the binning, usual
methods employ equiprobable bins so that $n_i^{\rm normal}$ is
constant; while others use a binning with a constant bin size (see
e.g. Ref.~\cite{eadie2006statistical} for more details on binning
strategies).

\subsubsection{Kolmogorov-Smirnov Test}

The Kolmogorov-Smirnov test (KS) quantifies the discrepancies between two
distributions by comparing their cumulative distribution functions (CDF). In
the case of a set of $N$ empirical data $X_i$, the CDF corresponds to
the fraction of data that is smaller than a certain value $x$ i.e.
\begin{equation}
 \label{eq:EmpiricalCDF}
 S_N(x) = \frac{1}{N}\sum_{i=1}^N\theta(x-x_i).
\end{equation}
The CDF of the normal distributions is given by
\begin{equation}
 \label{eq:NormalCDF}
 \Phi(x) = \frac1{\sqrt{2\pi}}\int_{-\infty}^x dt e^{-t^2/2}.
\end{equation}
The test statistic of this test is defined as the maximum value of the absolute 
difference between both cumulative distribution functions, that is
\begin{equation}
 T_{\rm KS} = \max_{-\infty < x < \infty}|S_N(x) - \Phi(x)|.
\end{equation}
Another advantage of the KS normality test is that a good approximation 
for the $p$-value exist and is given by
\begin{equation}
\label{eq:KSpvalue}
P_{\rm KS}(T_{\rm obs}) = 2 \sum_{j=1}^{\infty} (-1)^{j-1} e^{-2
\left[\left(\sqrt{N}+0.12+0.11/\sqrt{N}\right)jT_{\rm obs}\right]^2}.
\end{equation}
This approximation is sufficiently good for $N>4$ and approaches the 
actual value asymptotically as $N$ becomes larger.

For a normal distribution $x_i \in N[0,1] $ one has that for large $N$
and within $1\sigma$ confidence level,
\begin{equation}
S_N (x) = \phi(x) \pm  \frac1{\sqrt{N}} \sqrt{\phi(x)(1-\phi(x))} \, . 
\label{eq:meanCDF2}
\end{equation}

As in the Pearson test, a large value for $S_N(x)$
indicates larger discrepancies with the normal distribution and
therefore the null hypothesis will be rejected if $T_{\rm KS}$ is
larger than a certain critical values $T_{c,\rm KS}$. The critical
values depend on the sample size $N$ and level of significance
$\alpha$ with tables for common values of $\alpha$ and $N \leq 40$
readily available in the literature.  These same tables usually
include a fairly good approximation for $N > 40$, the more common case
in testing normality of residuals from a least squares fit, which we
reproduce here as a quick reference in table~\ref{tab:KSparameters}.
\begin{table}
 \caption{\label{tab:KSparameters} Large $N$ parametrization of the
   critical values for the Kolmogorov-Smirnov normality test $T_{\rm
     c} = a/\sqrt{N} $ at different levels of significance. This
   approximation is appropriate for $ N < 40$}
 \begin{ruledtabular}
 \begin{tabular*}{\columnwidth}{@{\extracolsep{\fill}}l c l l l c l}
   &$\alpha$& & & & a       \\
    \hline 
   & 0.01   & & & &1.63 \\
   & 0.02   & & & &1.52 \\
   & 0.05   & & & &1.36 \\
   & 0.10   & & & &1.22 \\
   & 0.20   & & & &1.07 \\
 \end{tabular*}
 \end{ruledtabular}
\end{table}

\subsubsection{QQ plot and tail sensitive test}

The idea of the Tail Sensitive (TS) test comes from the normal
quantile-quantile (QQ) plot, where one compares one to one the
empirical points $x_1 < \dots < x_N$ with the theoretical points
\begin{equation}
\frac{n}{N+1}= \int_{-\infty}^{x_n^{\rm th}} dx \frac{e^{-x^2/2}}{\sqrt{2\pi}}  \, , 
\label{eq:meanCDF3}
\end{equation}
If the empirical points follow a Normal distribution the QQ-plot
corresponds to the straight line rotated $45^0$, $x_n=x_n^{\rm th}$. A
good feature of this representation is that the tails are stretched
which makes it very suitable to analyze data selection where possible
outliers are discarded (see Sect.~\ref{sec:self-consistent}). The TS
test defines the test statistic as
\begin{eqnarray}
T_{\rm TS} &=& 2 
\min_i 
\left\{ \min \left[ B_{i,N+1-i}(z_i) , 1 - B_{i,N+1-i}(z_i) \right] 
\right\}, \nonumber \\
\end{eqnarray}
where the cumulative distribution function corresponds to the
regularized incomplete Beta-function and is defined as
\begin{equation}
B_{i,N+1-i} (z)=  \sum_{j=i}^{N}
\left( \begin{array}{c} N \\ j
     \end{array} 
\right) z^j (1-z)^{N-j},
\end{equation}
and the mapping between $N[0,1] \to U[0,1]$ 
\begin{equation}
z_i = \frac1{\sqrt{2\pi}} \int_{-\infty}^{x_i} dx e^{-x^2/2}, 
\end{equation}
has been introduced. In Ref.~\cite{Perez:2014kpa} 
the critical $T_c$ was tabulated  for $N \le 50$ and parametrized 
for $N> 50$ as 
\begin{equation}
T_c= \frac{a}{\sqrt{N}}+b, 
\end{equation}
for the usual significance levels of $\alpha=0.2,0.1,0.05,0.02, 0.01$.  

\section{Testing the UFD fits}
\label{sec:testing}

After the fitting process and the normality tests have been introduced, let
us now face the original UFD S0 wave fit from~\cite{GarciaMartin:2011cn} against them. 
As mentioned above, this is equivalent to check whether the statistical
assumptions implicit in the standard least squares $\chi^2$ fitting
procedure are satisfied.  Therefore, we will start computing the resulting
residuals of the fit, defined by Eq.~(\ref{residual-definition}) where
${\cal O}_i$ denotes the fitted observable, i.e. phase shift or
inelasticity data, and that should follow a
normal distribution within a given confidence level  

As we have pointed out in the previous section, 
a quantitative way to study the normal behavior of the residual
distribution is to analyze its central moments, which in the discrete
case are defined through,
\begin{equation}\label{mom}
  \mu_n=\frac{1}{N}\sum_{i=1}^N{(R_i-R_{\rm mean})^n}.
\end{equation}
In this way, we can compare whether they are compatible within
uncertainties with the expected moments of $N$ random data points
normally distributed, where again, $N$ is the total number of residual and
the moment error is given by the square of its variance. We can also
check whether this situation improves when we consider a normalized
residual distribution,
\begin{equation}\label{normlized-residual-definition}
\hat R_i=\frac{R_i-R_{\rm mean}}{\sqrt{V(R)}},
\end{equation}
where, $R_{\rm mean}=\Sigma_i^{N}{R_i}/N$ is the expected value of the
residual and $V(R)=\Sigma_i^N(R_i-R_{\rm mean})^2/(N-1)$ is its variance.
In this case, the central moments of the
distribution of the normalized residuals are given by:
\begin{equation}\label{normmom}
  \hat\mu_n=\frac{1}{N}\sum_{i=1}^N{(\hat R_i-\hat R_{\rm mean})^n}.
\end{equation}

For completeness, in Table~\ref{tab:eta-residuals}, we show the
experimental data considered, together with the fitted values and their
corresponding residuals.
Since the analysis of elastic $\pi\pi$ scattering in the $0^{++}$
channel was done separately in the low energy region and intermediate
region and also independently from the inelastic scattering data, we
have to pose the question on normality in a separate fashion.

\subsection{Low energy region}\label{subsec:low}

\begin{figure}
  \begin{center}
    \includegraphics[scale=0.35]{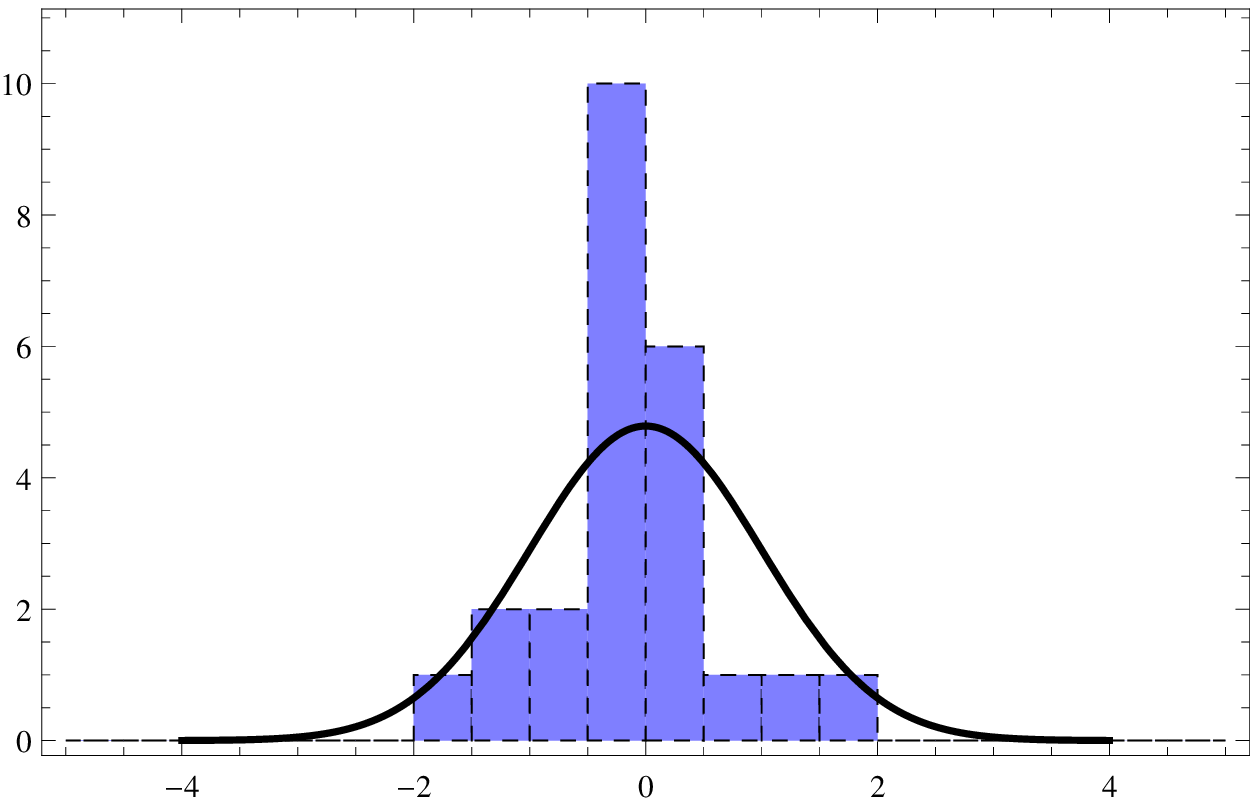}\includegraphics[scale=0.35]{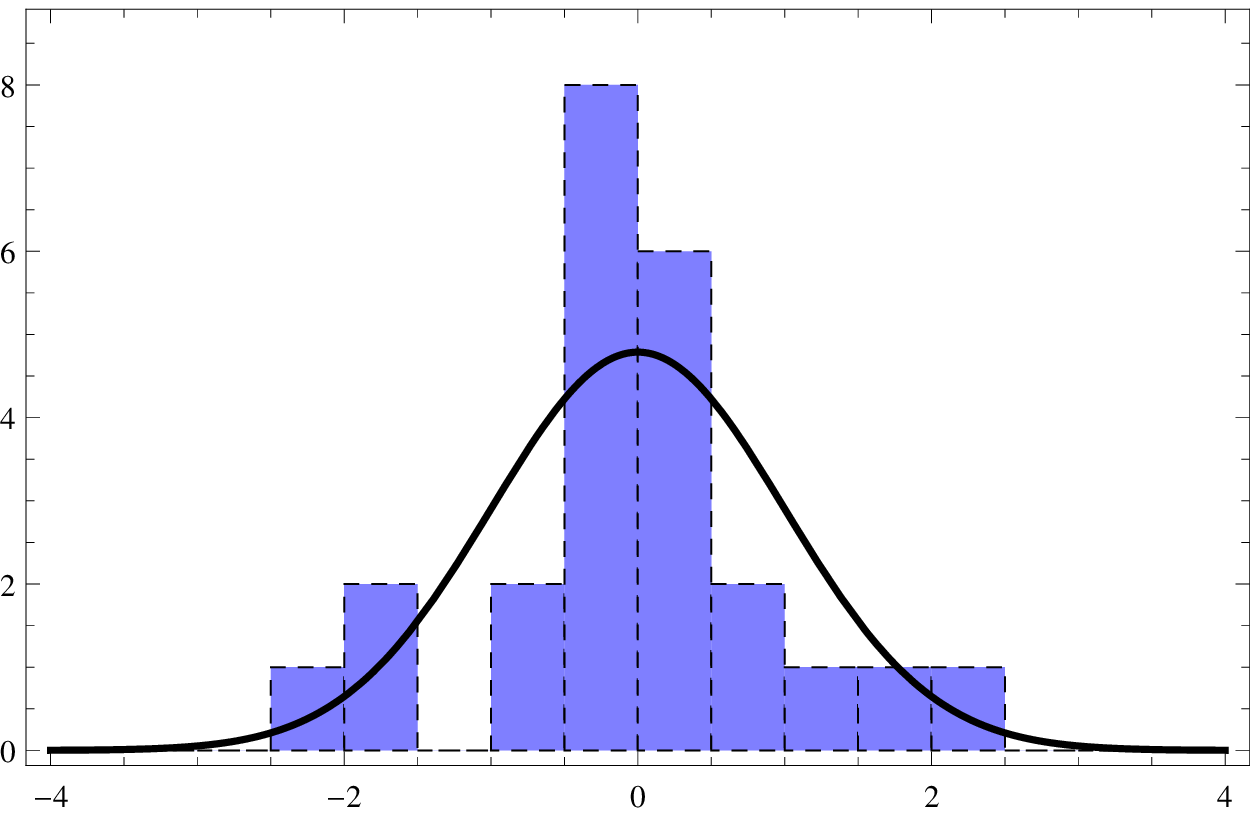}
    \includegraphics[scale=0.35]{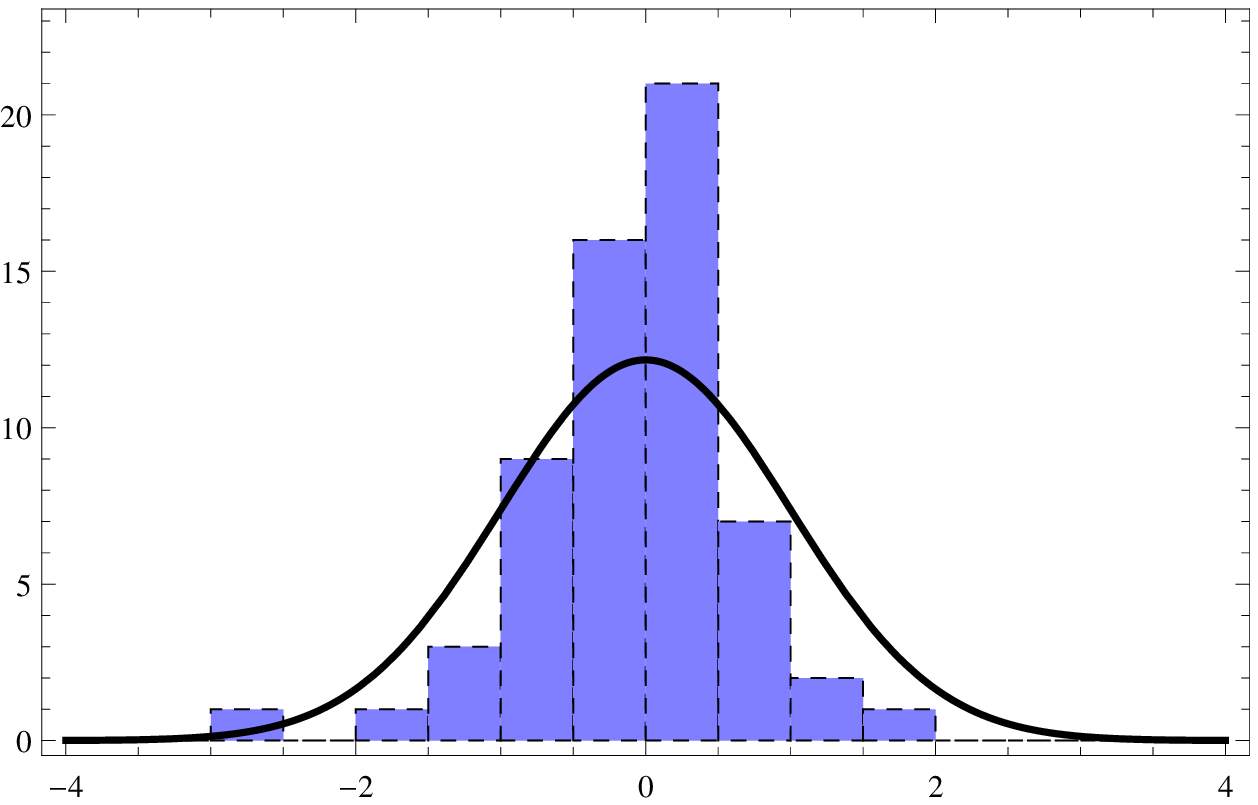}\includegraphics[scale=0.35]{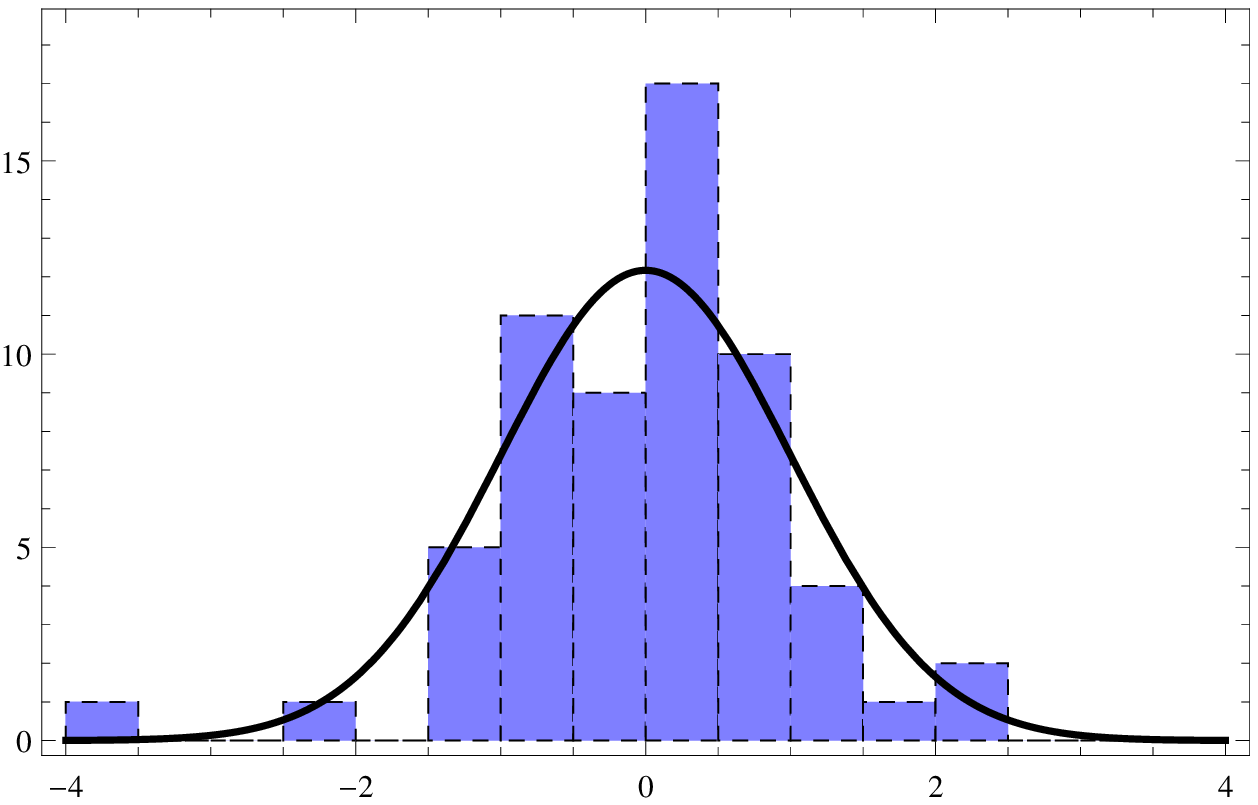}
    \includegraphics[scale=0.35]{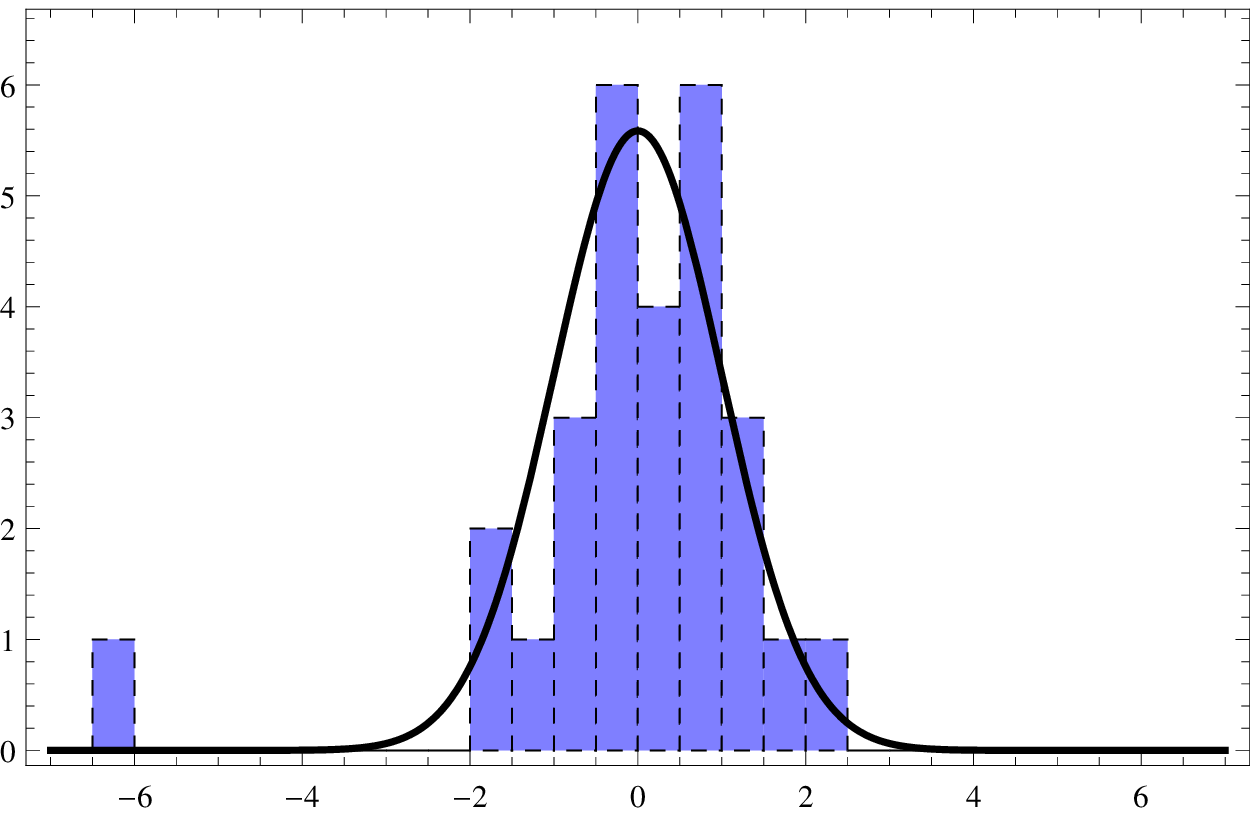}\includegraphics[scale=0.35]{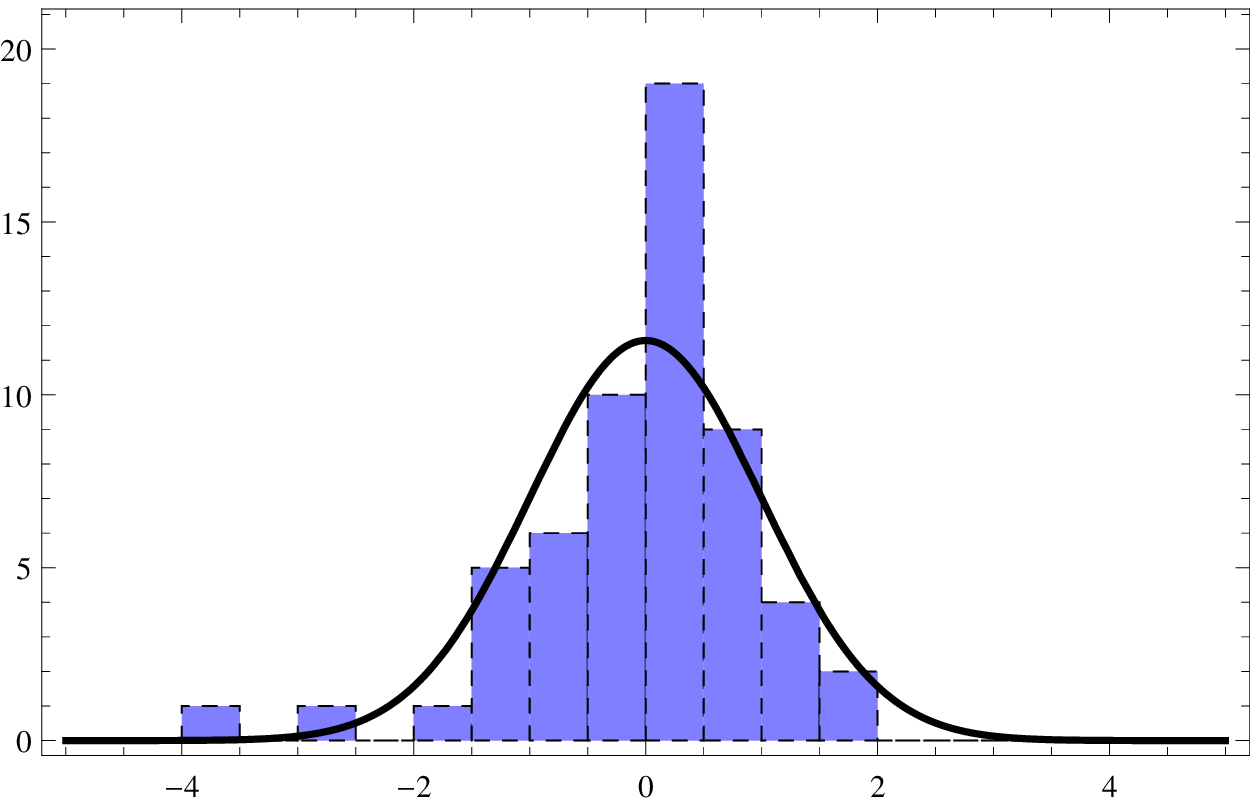}
  \end{center}
  \caption{Left: histograms of the resulting residuals after fitting
    the S0 wave data. Right: rescaled residual histogram.  Upper
    panels: Data described with conformal parametrization.  Middle
    panels: Data described by the polynomial parametrization.  Lower
    panels: Inelastic data. The black curve corresponds to an
    analytical normal distribution.
    \label{fig:initial}}
\end{figure}

In the upper left panel of Fig.~\ref{fig:initial}, we show for the
conformal low energy fit the distribution of the resulting residuals
given in Eq.~(\ref{residual-definition}) compared with the analytical
Gaussian distribution. We can conclude directly from this plot that it
is not justified to assume that they follow a normal
distribution. There is an excess of events in the origin.


In addition, in Table~\ref{tab:conf-mom-residuals} we show in both cases their
first moments. By definition, the first two moments are $\mu_0=1$,
$\mu_1=0$.  However, for the variance $\mu_2$, we obtain a tiny
deviation, which points out that the distribution of residuals cannot be considered Gaussian.  
Since the sample of low energy points ($N=30$) is rather small, 
for higher values of $n$, the Gaussian moment
uncertainties increase considerably, ensuring that the
residual moments are going to be satisfied within uncertainties.
\begin{table}[h]
  \centering
\begin{scriptsize}
  \begin{tabular}{cccccccc}
               & $\mu_0$  & $\mu_1$ &$\mu_2$ &$\mu_3$&$\mu_4$&$\mu_5$&$\mu_6$\\\hline
    $R_{\rm old}$& 1&0& 0.6&0.0&1.5&0.1&4.2\\
    $N(0,1)_{\rm old}$ & $1\pm0$&$0\pm0$&$1.0\pm0.3$&$0.0\pm0.5$&$2.8\pm2.1$&$0\pm 6$&$14\pm 30$\\
    $R_{\rm new}$& 1&0& 0.9&-0.1&2.1&-0.2&7\\
    $N(0,1)_{\rm new}$ & $1\pm0$&$0\pm0$&$1.0\pm0.3$&$0.0\pm0.5$&$2.8\pm1.9$&$-0.1\pm 5$&$14\pm 19$\\\hline\\
    & $\hat\mu_0$  & $\hat\mu_1$ &$\hat\mu_2$ &$\hat\mu_3$&$\hat\mu_4$&$\hat\mu_5$&$\hat\mu_6$\\\hline
    $\hat R_{\rm old}$ &1&0& 1&0&3.5&0.4&16\\
    $\hat N(0,1)_{\rm old}$ & $1\pm0$&$0\pm0$&$1\pm0$&$0.0\pm0.4$&$2.5\pm0.6$&$0\pm 3$&$10\pm 7$\\
    $\hat R_{\rm new}$& 1&0& 1&-0.1&2.6&-0.3&9\\
    $\hat N(0,1)_{\rm new}$ & $1\pm0$&$0\pm0$&$1\pm0$&$0.0\pm0.4$&$2.6\pm0.7$&$0\pm 3$&$11\pm 7$\\\hline
  \end{tabular}
\end{scriptsize}
  \caption{Central moments of the distribution of residuals obtained for
    the low energy conformal fit versus the moments of N random points normally
    distributed.  
    The subscripts ``old'' and ``new'' denote respectively the original UFD and the 
    UFD for the new selection of data performed in Section~\ref{sec:new-UFD}.
    $\mu_0$ and $\mu_1$ are by definition 1 and 0
    respectively. However for $\mu_2$, very small deviations appear for the original UFD. 
    For higher values of $n$, they are always compatible due to huge
    uncertainties coming from a small sample of points (N=30). 
    In contrast, for the new selection of data all moments are compatible 
    with the expected values of a gaussian distribution.
    The last four rows contain the central moments of the normalized
    residual distribution obtained in the conformal fit versus the
    moments of N normalized random points normally distributed.}
  \label{tab:conf-mom-residuals}
\end{table}

In the upper right panel of Fig.~\ref{fig:initial}, we show the
normalized residual histogram, where we can see, that, despite an
important improvement, it does not correspond to a normal
distribution.  This can be again checked by comparing the last two
rows in Table~\ref{tab:conf-mom-residuals}, where we show the central
moments of the normalized residual distribution and those of a
rescaled Gaussian distribution of N random points. The first
deviations occur for $\hat\mu_4$.  Therefore, we can conclude that
although only slightly, the low energy S0 wave phase shift UFD fit
from~\cite{GarciaMartin:2011cn} violates the residual normality test.
However, this small violation anticipates tiny corrections to the
results obtained in~\cite{GarciaMartin:2011cn}.

\subsection{Intermediate energy}\label{subsec:inter}

As we did for the low energy case, 
we will start analyzing the normal behavior of the intermediate energy S0 wave phase shift fit 
by studying the behavior of its residuals. 
Their normalized and rescaled distributions are plotted in the middle panel of 
Fig.~\ref{fig:initial}, and point out again a excess of residual at the origin,  
and consequently a non-gaussian behavior.
As we proceeded for the previous fit, in order to quantify this
deviation, we compute again the moments
for the normal and rescales residual distributions. 
The results are given in Table~\ref{tab:pol-mom-residuals}.
\begin{table}[h]
  \centering
  \begin{scriptsize}
  \begin{tabular}{lccccccc}
             & $\mu_0$  & $\mu_1$ &$\mu_2$ &$\mu_3$&$\mu_4$&$\mu_5$&$\mu_6$\\\hline
    $R_{\rm old}$& 1&0& 0.5&-0.3&1.3&-2&7\\
    $N(0,1)_{\rm old}$ & $1\pm0$&$0\pm0$&$1.0\pm0.2$&$0.0\pm0.3$&$2.9\pm1.2$&$0\pm 3$&$14\pm 12$\\
    $R_{\rm new}$& 1&0& 0.8&0.0&1.3&-0.2&4\\
    $N(0,1_{\rm new})$ & $1\pm0$&$0\pm0$&$1.0\pm0.2$&$0.0\pm0.3$&$2.8\pm1.2$&$-0.2\pm 3$&$14\pm 11$\\\hline\\

    & $\hat\mu_0$  & $\hat\mu_1$ &$\hat\mu_2$ &$\hat\mu_3$&$\hat\mu_4$&$\hat\mu_5$&$\hat\mu_6$\\\hline
    $\hat R_{\rm old}$& 1&0& 1&-0.8&5.2&-12&53\\
    $\hat N(0,1_{\rm old})$ & $1\pm0$&$0\pm0$&$1\pm0$&$0.0\pm0.3$&$2.8\pm0.5$&$0\pm 3$&$13\pm 8$\\
    $\hat R_{\rm new}$& 1&0& 1&0.1&2.4&0.6&8\\
    $\hat N(0,1)_{\rm new}$ & $1\pm0$&$0\pm0$&$1\pm0$&$0.0\pm0.3$&$2.7\pm0.6$&$0\pm 3$&$12\pm 8$\\\hline
  \end{tabular}
  \end{scriptsize}
  \caption{Central non- and normalized moments for the residuals of
    the intermediate energy polynomial parametrization for the original and new UFD of Section~\ref{sec:new-UFD}.}
  \label{tab:pol-mom-residuals}
\end{table}

Since for the intermediate energy region the sample of points is bigger ($N=61$),
the expected errors are smaller and the residual constraint becomes this time much more stringent. 
For the non-rescaled case we can see small deviation for $\mu_2$ and $\mu_4$, 
proving again a non Gaussian behavior. For the rescaled case bigger deviations occur for $n>3$. 

\subsection{Inelastic data}

In the case of the S0 wave inelasticity,
the non- and normalized distribution of the residuals 
are plotted in the lower panel of Fig.~\ref{fig:initial}.  
In this case, the main problem does not come from an excess of events at the origin, 
but for some values which are far away from it, in contradiction with a gaussian behavior.
The central moments of both distributions are given in Tab.~\ref{tab:eta-mom-residuals} 
and they show that, despite we are working with a small sample ($N=28$) and the expected errors are big, 
there are severe deviations, some of them of more than an order of magnitude. 
\begin{table}[h]
  \centering
  \begin{scriptsize}
  \begin{tabular}{lccccccc}
             & $\mu_0$  & $\mu_1$ &$\mu_2$ &$\mu_3$&$\mu_4$&$\mu_5$&$\mu_6$\\\hline
    $R_{\rm old}$& 1&0& 2.2&-7.9&54&-318&1981\\
    $N(0,1)_{\rm old}$ & $1\pm0$&$0\pm0$&$1.0\pm0.2$&$0.0\pm0.4$&$3.0\pm1.3$&$0\pm 3$&$14\pm 12$\\
    $R_{\rm new}$& 1&0& 0.8&0.0&1.9&0.0&6\\
    $N(0,1_{\rm new})$ & $1\pm0$&$0\pm0$&$1.0\pm0.2$&$0.0\pm0.3$&$2.8\pm1.2$&$0\pm 3$&$14\pm 11$\\\hline\\
    & $\hat\mu_0$  & $\hat\mu_1$ &$\hat\mu_2$ &$\hat\mu_3$&$\hat\mu_4$&$\hat\mu_5$&$\hat\mu_6$\\\hline
    $\hat R_{\rm old}$& 1&0& 1&-0.8&5.2&-12&53\\
    $\hat N(0,1)_{\rm old}$ & $1\pm0$&$0\pm0$&$1\pm0$&$0.0\pm0.3$&$2.8\pm0.5$&$0\pm 3$&$13\pm 7$\\
    $\hat R_{\rm new}$& 1&0& 1&0.1&2.4&0.6&8\\
    $\hat N(0,1)_{\rm new}$ & $1\pm0$&$0\pm0$&$1\pm0$&$0.0\pm0.3$&$2.8\pm0.5$&$0\pm 3$&$13\pm 7$\\\hline
  \end{tabular}
  \end{scriptsize}
  \caption{Central moments of the residual distribution obtained for
    the S0 wave inelasticity for the original (old) and new selection of data (new) (Section~\ref{sec:new-UFD}). 
    In the case of the original UFD, there are very severe deviation with respect to a normal distribution.
    On the contrary, for the new selection of data, the agreement is remarkable.}
  \label{tab:eta-mom-residuals}
\end{table}

\subsection{Quantitative checks}

\begin{table}[h]
 \caption{\label{tab:PearsontestResults} Results of the Pearson
   normality test of the residuals obtained by fitting the different
   data sets with the conformal, polynomial and inelastic
   parametrizations. The results of the test of the scaled residuals
   for every case is shown below the corresponding line. The critical
   value $T_c$ corresponds to a significance level of $\alpha=0.05$.}
 \begin{ruledtabular}
 \begin{tabular*}{\columnwidth}{@{\extracolsep{\fill}} c c c D{.}{.}{2.3} D{.}{.}{3.2} D{.}{.}{1.9}}
   Database  & Fit  &$N$ &  \multicolumn{1}{c}{$T_{c}$} & \multicolumn{1}{c}{$T_{\rm obs}$} & \multicolumn{1}{c}{$p$-value}  \\
  \hline  
   Conformal      & UFD & $30$ & 14.07 & 14.80 & 0.039   \\
                  &     &      &       & 10.43 & 0.021   \\
   Conformal new  & UFD & $25$ & 14.07 &  5.40 & 0.611   \\
                  &     &      &       &  5.40 & 0.611   \\
   Polynomial     & UFD & $61$ & 18.31 & 21.05 & 0.021   \\
                  &     &      &       &  6.98 & 0.727   \\
   Polynomial new & UFD & $52$ & 16.92 &  4.15 & 0.901   \\
                  &     &      &       &  3.00 & 0.964   \\
   Inelastic      & UFD & $28$ & 14.07 &  5.14 & 0.642   \\
                  &     &      &       &  6.71 & 0.573   \\
   Inelastic new  & UFD & $27$ & 14.07 &  2.93 & 0.892   \\
                  &     &      &       &  2.93 & 0.892   \\
 \end{tabular*}
 \end{ruledtabular}
\end{table}
\begin{table}[h]
 \caption{\label{tab:KStestResults} Same as
 table~\ref{tab:PearsontestResults} for the Kolmogorov-Smirnov test}
 \begin{ruledtabular}
 \begin{tabular*}{\columnwidth}{@{\extracolsep{\fill}} c c c D{.}{.}{1.3} D{.}{.}{1.3} D{.}{.}{1.9}}
   Database & Fit  & $N$ &  \multicolumn{1}{c}{$T_{c}$} & \multicolumn{1}{c}{$T_{\rm obs}$} & \multicolumn{1}{c}{$p$-value}  \\
  \hline  
  \hline  
   Conformal      & UFD & $30$ & 0.242 & 0.253 & 0.036   \\
                  & UFD &      &       & 0.216 & 0.105   \\
   Conformal new  & UFD & $25$ & 0.264 & 0.158 & 0.513   \\
                  & UFD &      &       & 0.157 & 0.520   \\
   Polynomial     & UFD & $61$ & 0.174 & 0.170 & 0.053   \\
                  & UFD &      &       & 0.097 & 0.582   \\
   Polynomial new & UFD & $52$ & 0.189 & 0.082 & 0.846   \\
                  & UFD &      &       & 0.058 & 0.991   \\
   Inelastic      & UFD & $28$ & 0.250 & 0.118 & 0.785   \\
                  & UFD &      &       & 0.186 & 0.254   \\
   Inelastic new  & UFD & $27$ & 0.254 & 0.140 & 0.611   \\
                  & UFD &      &       & 0.129 & 0.715   \\
 \end{tabular*}
 \end{ruledtabular}
\end{table}
\begin{table}[h]
 \caption{\label{tab:TStestResults} Same as
 table~\ref{tab:PearsontestResults} for the tail sensitive test}
 \begin{ruledtabular}
 \begin{tabular*}{\columnwidth}{@{\extracolsep{\fill}} c c c D{.}{.}{1.5} D{.}{.}{1.9} D{.}{.}{3.4}}
   Database & Fit & $N$ &  \multicolumn{1}{c}{$T_{c}$} & \multicolumn{1}{c}{$T_{\rm obs}$} & \multicolumn{1}{c}{$p$-value}  \\
  \hline  
   Conformal      & UFD & $30$ & 0.0039 & 0.0049 & 0.061   \\
                  & UFD &      &        & 0.0155 & 0.164   \\
   Conformal new  & UFD & $25$ & 0.0042 & 0.1662 & 0.794   \\
                  & UFD &      &        & 0.1686 & 0.799   \\
   Polynomial     & UFD & $61$ & 0.0027 & 0.0010 & 0.020   \\
                  & UFD &      &        & 0.0081 & 0.132   \\
   Polynomial new & UFD & $52$ & 0.0029 & 0.0912 & 0.681   \\
                  & UFD &      &        & 0.3101 & 0.988   \\
   Inelastic      & UFD & $28$ & 0.0040 & 1.2\times10^{-8}  & 0.0002  \\
                  & UFD &      &        & 0.001  & 0.016   \\
   Inelastic new  & UFD & $27$ & 0.0040 & 0.1505 & 0.769  \\
                  & UFD &      &        & 0.1913 & 0.851   \\
 \end{tabular*}
 \end{ruledtabular}
\end{table}

\begin{figure*}[htbp]
\begin{center}
\epsfig{figure=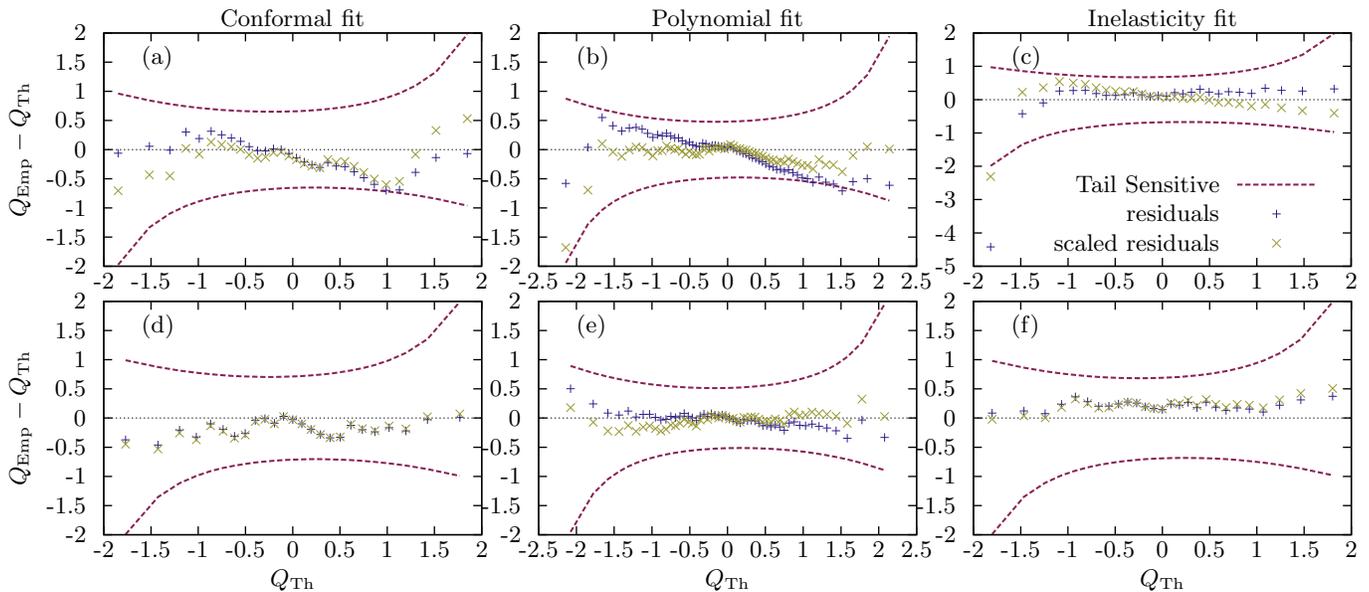,width=\linewidth}  
\end{center}   
\caption{(Color online) Rotated Quantile-Quantile plots for residuals
  of different fits.  Left panels: low energy conformal fits. Middle
  panels: intermediate energy polynomial fits.  Right panels: high
  energy inelasticity fits. Upper panels: fit to the original
  database.  Lower panels: fit to the newly selected data. The
  resulting residuals are marked as blue vertical crosses and the
  scaled ones as yellow diagonal crosses.  The $95 \%$ confidence
  bands from the TS normality test are also shown (dashed red line).}
\label{fig:QQTSbands}
\end{figure*}

On a more quantitative level, we complement the previous results with
Tables~\ref{tab:PearsontestResults},\ref{tab:KStestResults}
\ref{tab:TStestResults} for the Pearson, KS and TS tests respectively.
It is interesting to note that for the Pearson and KS tests the
p-value is very high for the Inelastic fit, whereas the TS gives a too
low value.  This is mainly due to the outlier with $R_i=-6.24$ sitting
at the tail of the distribution. We also note that for these small data points
values ($N$), the power of the test (probability of not giving a false
positive) is low. Finally, QQ-plots are shown in
Fig.~\ref{fig:QQTSbands}. All this reinforces the lack of normality of
residuals.

\section{New selection of data}
\label{sec:new-UFD}

The results presented in Section~\ref{sec:testing} highlight that the UFD
fits to the $\pi\pi$ scattering amplitude in the $0^{++}$ channel from~\cite{GarciaMartin:2011cn} 
do not satisfy {\it a posteriori} the statistical assumptions implicit in the
$\chi^2$ minimization. This inconsistency in the $\chi^2$ fit can be
due to either the presence of mutually incompatible data or a biased
choice of the fitting curve.  We will explore the consequences that a
new selection of data implies regarding normality of residuals.

\subsection{New selection of low energy data}
As we have seen before, there are three different sets of
experiments included in the low energy region.  
The NA48/2 set~\cite{ULTIMONA48} corresponds to the
newest data on $K_{\ell4}$ decays and incorporates a rigorous
statistical and systematic error analysis, so there is no reason to
modify or alter this set of data.  As we have commented before, the
PY05 average selection corresponds to an average of the different
experimental solutions that passed a consistency test with FDR and
other sum rules. The large uncertainties assigned covered the
difference between the different data sets, and at some point, they
can be considered arbitrary. Since, as we can see in
upper Fig.~\ref{fig:initial}, there is an excess of residuals at
the origin, it is reasonable to assume that they may be
overestimated. The last set of data considered in the low energy fit corresponds
to old data on $K_{\ell4}$ decays from~\cite{Rosselet:1976pu}, whose
precision cannot be consider as the same level than the recent NA48/2 analysis. 
Therefore, it makes sense to get rid of those old data points on $K_{\ell4}$ decays
which lie within uncertainties in the same energy bin than a NA48/2 value.  
Thus, in order to improve the Gaussian check of the fit, we divide by two
the uncertainty of the PY05 data average and get rid of the old K
decay corresponding to the following values of $\sqrt{s}$, \{289, 317,
324, 340, 367\}.  The comparison of the tiny differences between the
old and the new fit is plotted in Fig.~\ref{fig:low-comparison}.  Its
corresponding parameters are given in Table~\ref{tab:S0parameters}.
\begin{figure}
  \begin{center}
  \hspace{0cm} \includegraphics[scale=0.25]{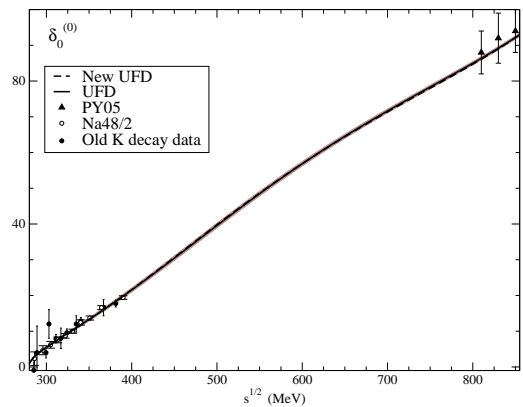}
  \end{center}
  \caption{Old versus new UFD at low energies, the dark band covers
    the uncertainties of the old parametrization.
    \label{fig:low-comparison}}
\end{figure}

Finally, in the upper panel of Fig.~\ref{fig:new-fit}, we plot the residual
distribution for the new fit. Despite there are still some deviations,
there is a clear improvement, which can be again checked by computing
the central moments of the distribution. They are given in
Table~\ref{tab:conf-mom-residuals}, showing a complete agreement between them
and the expected values of N random points normally distributed.

\begin{figure}
  \begin{center}
    \includegraphics[scale=0.35]{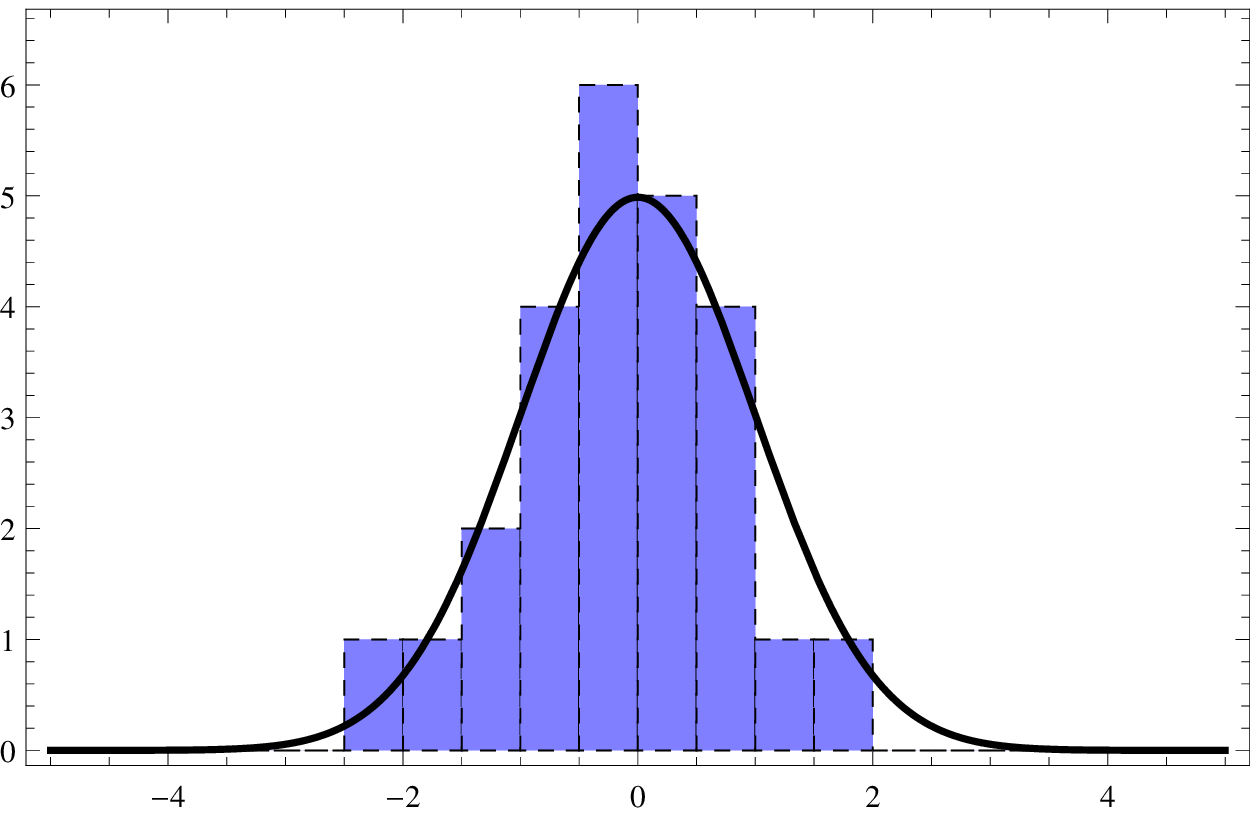}\includegraphics[scale=0.35]{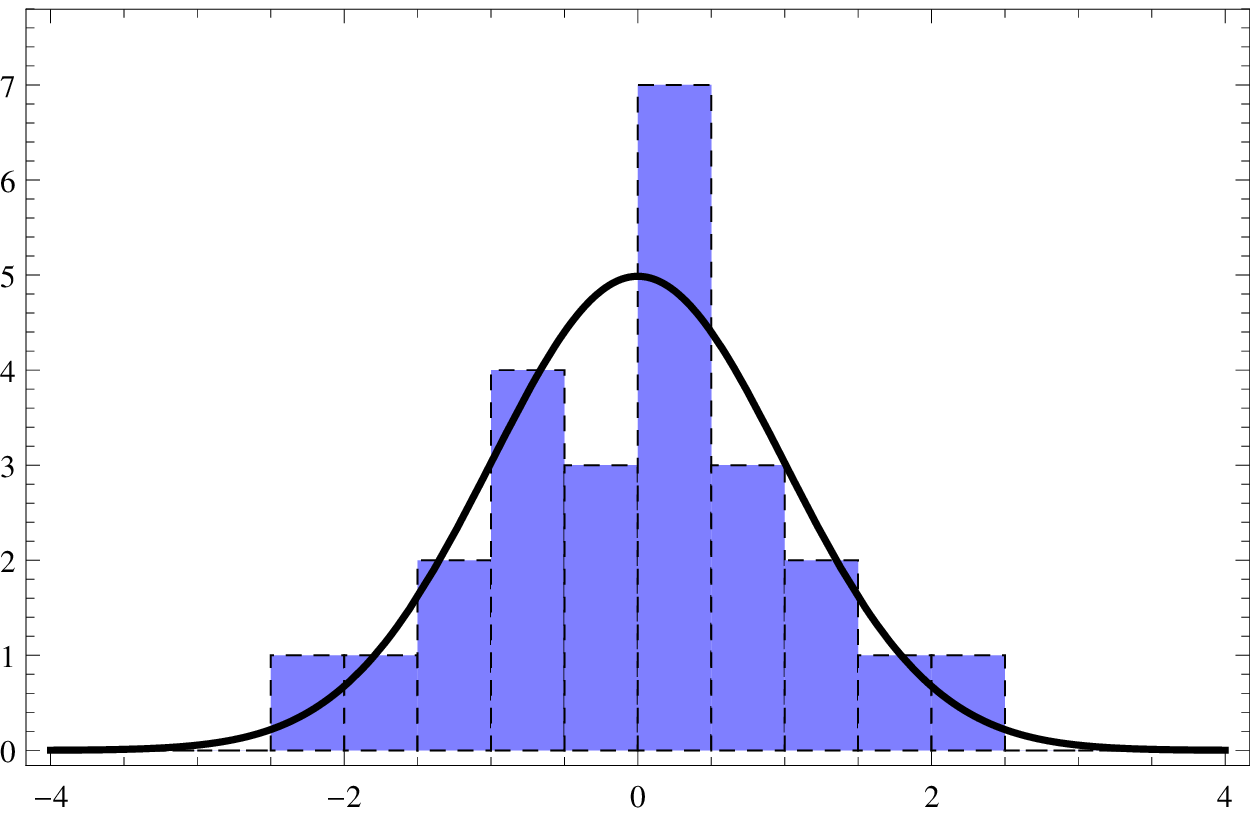}
    \includegraphics[scale=0.35]{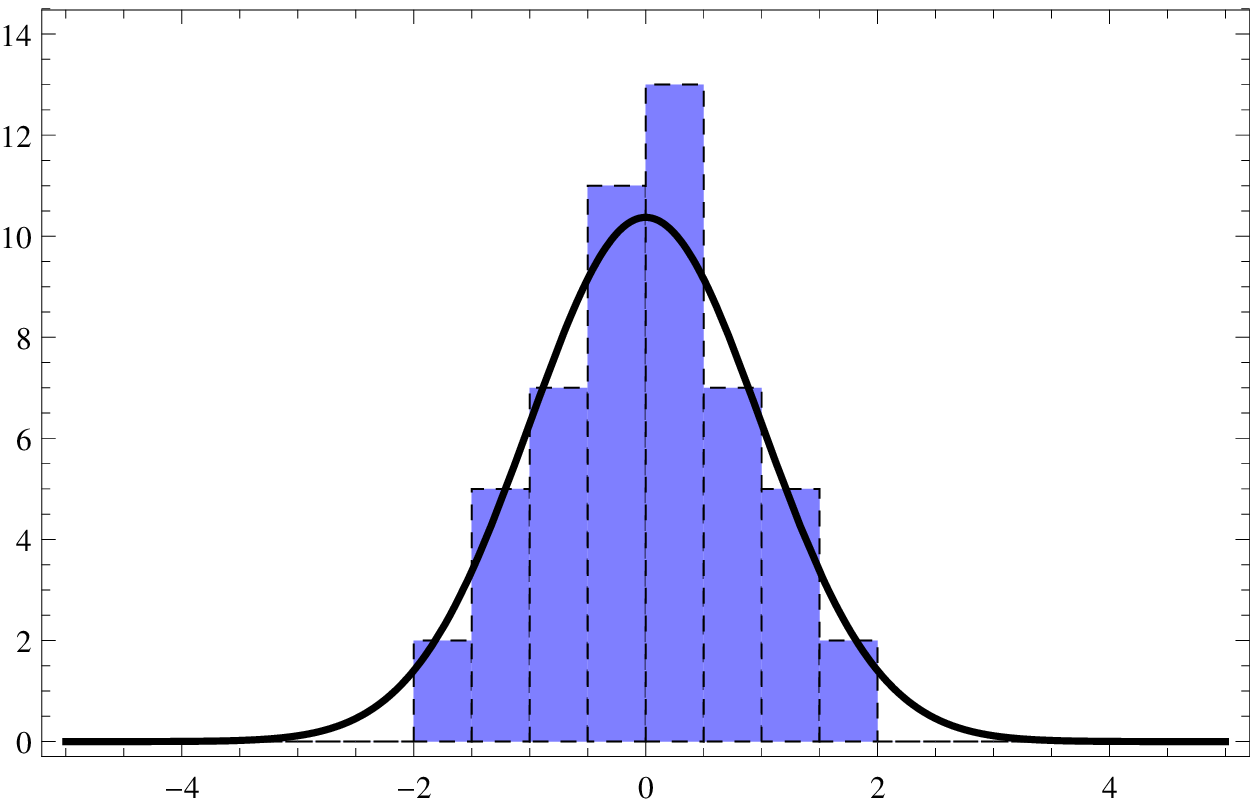}\includegraphics[scale=0.35]{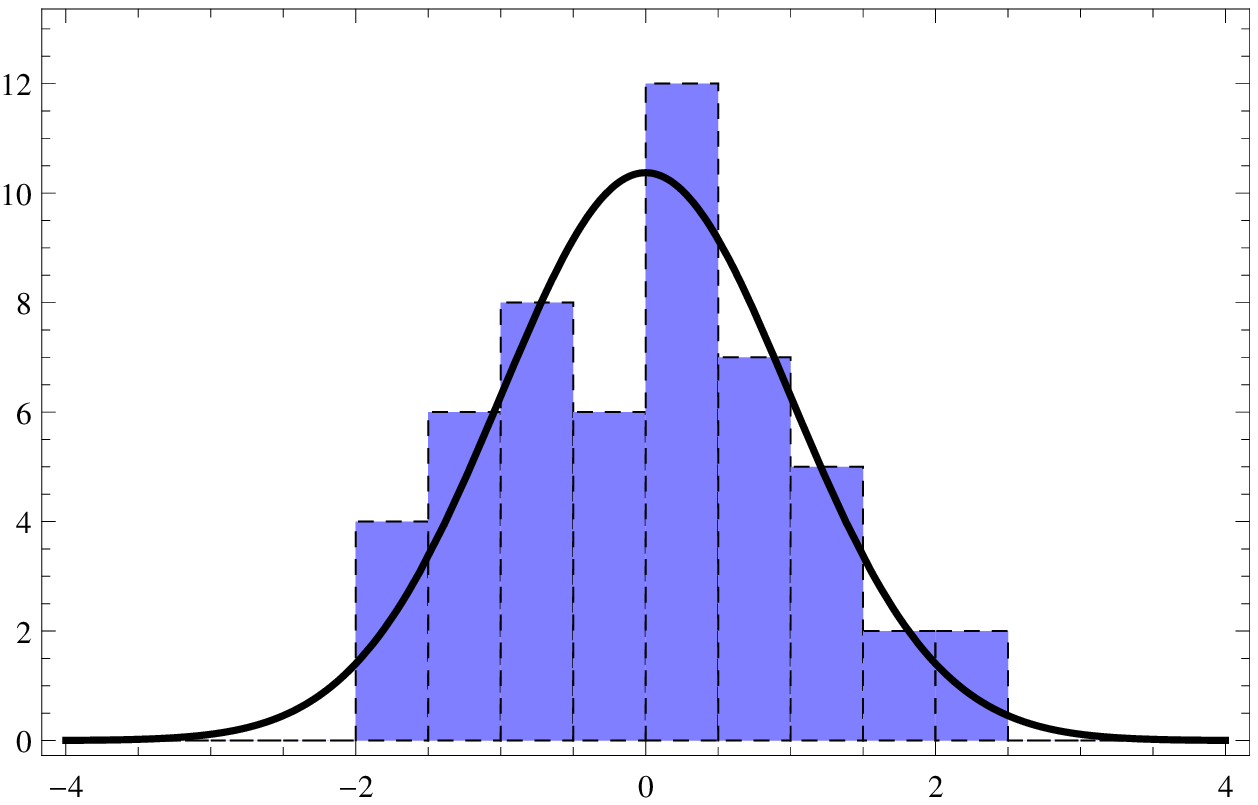}
    \includegraphics[scale=0.35]{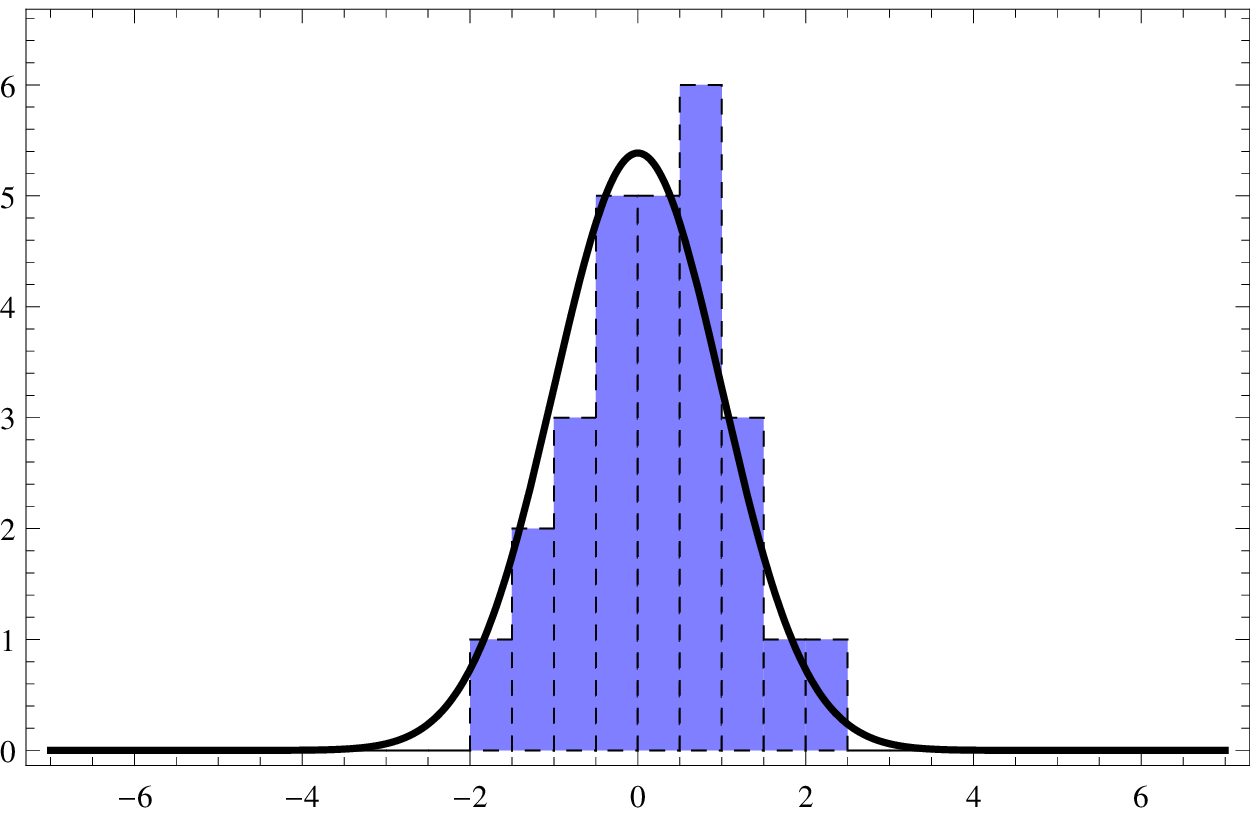}\includegraphics[scale=0.35]{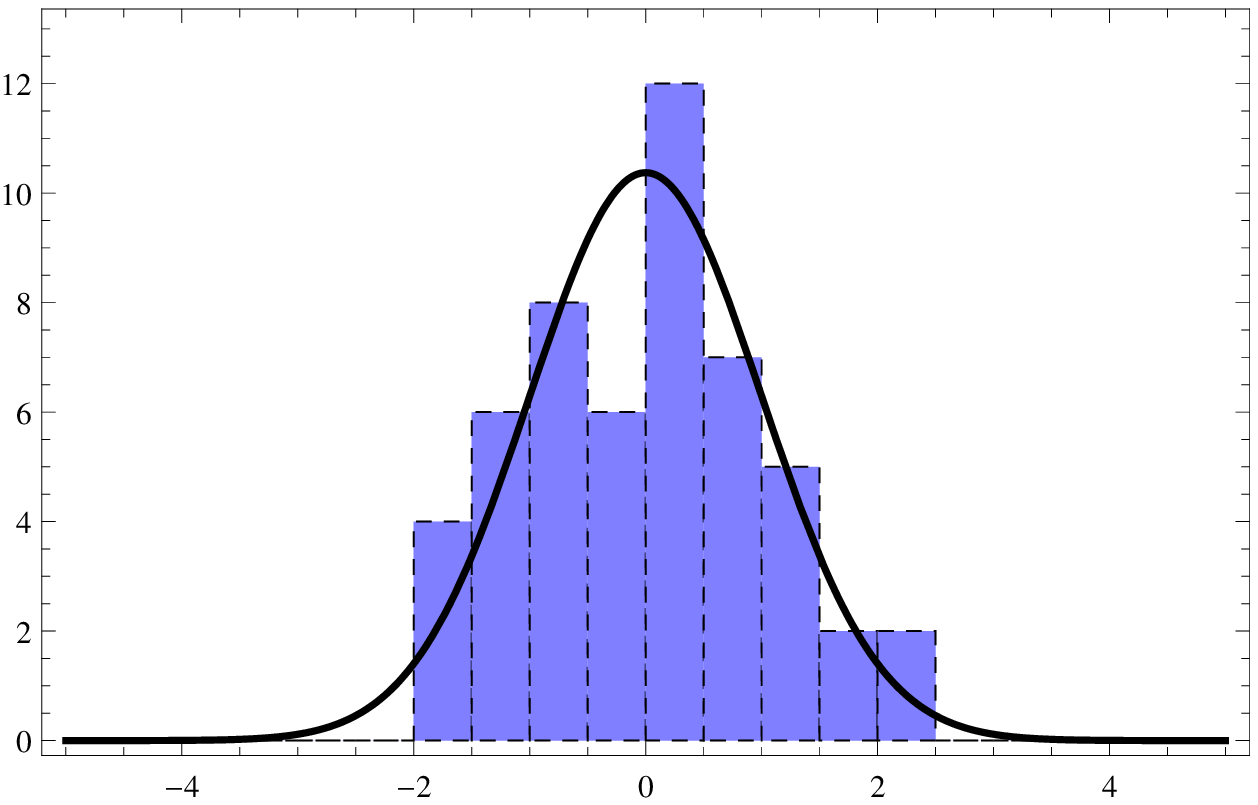}
  \end{center}  \caption{Same as Fig.~\ref{fig:initial} for the new data selection.  
\label{fig:new-fit}}
\end{figure}

In this was, by imposing the Gaussian check in the data selection, we
are able to constraint part of the arbitrariness assumed in the data
selection, namely, the systematic uncertainty taken for the PY05
average and the old K decay data points considered.

\subsection{New selection of intermediate energy data}
On the one hand, as we can can see in
middle Fig.~\ref{fig:initial}, there is again a excess of residuals
in the central region, i.e. for $R\sim0$, which could be considered as
a signal of overestimated uncertainties.  On the other hand, the
uncertainties of the PY05 average data, the systematic error added to
Hyams and the errors given by Kaminski et al. can be considered at
some point arbitrary.  Therefore, in order to improve the normal
behavior of the residual distribution, we reduce by a factor 1/2 the
uncertainties of the PY05 average, we sum $3^\circ$ instead of
$5^\circ$ as systematic uncertainty to the data from Hyams, and
rescale by a factor of 3/4 the uncertainties of Kaminski data.  
In addition, since there are several Kaminski data points which are compatible with those of
Grayer, Hyams or PY05 but with higher uncertainties, we get rid of the
Kaminski data point at the following energies
\{930,970,1010,1050,1210,1390\}.  The normalized and rescaled
distribution of the residuals for the new data selection are plotted
in middle Fig.~\ref{fig:new-fit}, whereas the central moments are given in
Table~\ref{tab:pol-mom-residuals}. Despite there are still some deviations, the
improvement is clear.  The value of the parameters corresponding to
the new polynomial fit are given in Table~\ref{tab:S0parameters}, and
the comparison between the new and old polynomial curve is plotted in
Fig.~\ref{fig:pol-comparison}.
\begin{figure}
  \begin{center}
  \hspace{0cm} \includegraphics[scale=0.25]{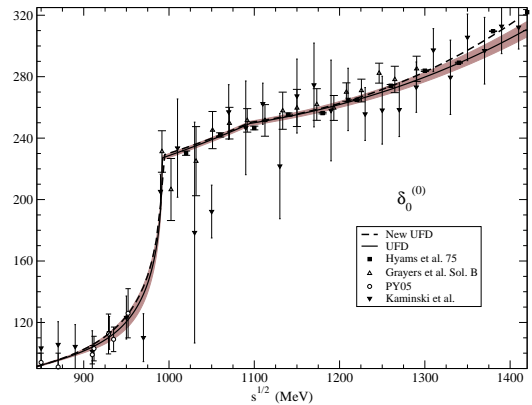}
  \end{center}
  \caption{Old versus new UFD for the polynomial parametrization, the dark band covers the uncertainties of the old parametrization.
    \label{fig:pol-comparison}}
\end{figure}

\subsection{New selection of  inelasticity data}
In this case, as we can see in the histograms depicted in
the lower panel of Fig.~\ref{fig:initial}, rather than by overestimated errors,
the problem comes from incompatible data points, and in particular for
the Hyams data point at $\sqrt s=994$ MeV.  If we simply get rid of
this point, the new fit improves striking the normal behavior as
we can see in the lower panel of Fig.~\ref{fig:new-fit}.  The central moments of the
new fit are given in Table~\ref{tab:eta-mom-residuals}, and show a perfect
agreement.  The new inelasticity parameters are given in
Table~\ref{tab:S0parameters}, and the difference between the old and
new parametrizations are plotted in Fig.~\ref{fig:eta-comp}.
\begin{figure}
  \begin{center}
  \hspace{0cm} \includegraphics[scale=0.25]{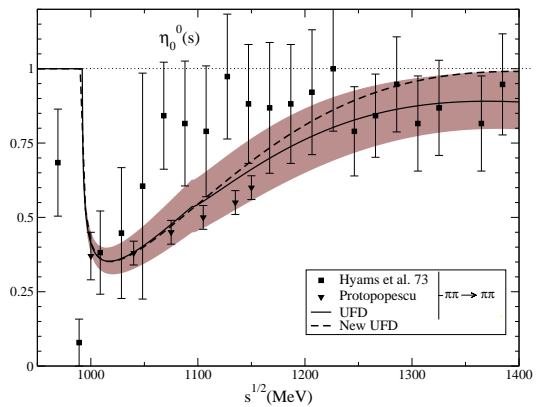}
  \end{center}
  \caption{Old versus new UFD for the inelasticity, the dark band covers the uncertainties of the old parametrization.
    \label{fig:eta-comp}}
\end{figure}

\begin{table}[h]
  \centering
  \begin{scriptsize}
  \begin{tabular}{ccc}
S0 wave   & UFD &New UFD\\
\hline
$ B_0$ &$7.26\pm0.23$ &$7.38\pm0.23$ \\
   $B_1$& $-25.3\pm0.5$&$-24.5\pm0.5$\\
 $B_{2}$&$-33.1\pm1.2$&$-35.4\pm1.2$\\
 $B_{3}$&$-26.6\pm2.3$&$-34.7\pm2.7$\\
 $z_0$ & $M_\pi$& $M_\pi$\\
$\chi^2/\nu$ & $0.645$   & $1.111$ \\
$1-\sqrt{2/\nu}$ & $0.723$ & $0.691$  \\
$1+\sqrt{2/\nu}$ & $1.277$ & $1.309$  \\
\hline
$d_0$ & $(227.1\pm1.3)^\circ$& $(228.8\pm1.3)^\circ$\\
$c$ & $(-660\pm  290)^\circ$& $(-466\pm  294)^\circ$\\
$B$ &  $(94.0\pm 2.3)^\circ$& $(85.1\pm 2.3)^\circ$ \\
$C$ & $(40.4\pm 2.9)^\circ$& $(60.7\pm 2.9)^\circ$ \\
$D$ & $(-86.9\pm 4.0)^\circ$& $(-92.3\pm 4.0)^\circ$ \\
$\chi^2/\nu$ & $0.552$   & $0.761$ \\
$1-\sqrt{2/\nu}$ & $0.811$ & $0.794$  \\
$1+\sqrt{2/\nu}$ & $1.189$ & $1.206$  \\
\hline
$\tilde\epsilon_1$ & $4.7\pm0.2$  & $4.6\pm0.1$  \\
$\tilde\epsilon_2$ & $-15.0\pm0.8$  & $-14.6\pm0.4$  \\
$\tilde\epsilon_3$ & $4.7\pm2.6$ & $3.9\pm4.4$ \\
$\tilde\epsilon_4$ & $0.38\pm0.34$ & $0.37\pm0.20$ \\
$\chi^2/\nu$ &  $2.668$  & $1.039$ \\
$1-\sqrt{2/\nu}$ & $0.711$ & $0.705$  \\
$1+\sqrt{2/\nu}$ & $1.289$ & $1.294$  \\
\hline
  \end{tabular}
  \end{scriptsize}
  \caption{S0 wave parameters for the old and new UFD sets, indicating
    also the number of fitted data $N$, the corresponding $\chi^2/\nu$
    and expected value boundaries $1\pm \sqrt{2/\nu}$.  The first set
    corresponds to the low energy parametrization, $\sqrt{s}\leq
    0.85\,\mathrm{GeV}$, the second set to the intermediate energy
    conformal fit and the last one to the inelastic parametrization up
    to $\sqrt{s}=1.42\,\mathrm{GeV}$.}
  \label{tab:S0parameters}
\end{table}

\subsection{Resonance pole parameters}
As it has been pointed out in the introduction, one of the main
achievements of the dispersive $\pi\pi$ analyses based on Roy and
Roy-like
equations~\cite{Ananthanarayan:2000ht,Colangelo:2001df,GarciaMartin:2011cn}
has been the precise and unbiased determination of the lowest scalar
resonance states~\cite{Caprini:2005zr,GarciaMartin:2011jx}, namely,
the $f_0(500)$ and $f_0(980)$.  In particular,
in~\cite{GarciaMartin:2011jx}, the CFD parametrizations
of~\cite{GarciaMartin:2011cn} were used as input for the analytic
extrapolation to the complex plane of the dispersive once- and
twice-subtracted Roy equation, and thus, allowing to look for the
lowest-lying poles on the second Riemann sheet.  Therefore, it is
relevant to check the effect of the new selection of data on the
scalar pole determinations.  In Table~\ref{tab:poles}, we show the
pole positions of the $f_0(500)$ and $f_0(980)$ resonances for the old
and new UFD, obtained from the twice-subtracted Roy equations. For
completeness, we also show the results for the $\rho(770)$ resonance.
Note that the values of the $\rho$ pole depend on the new S0 wave data
selection through the dispersive integral.  The small differences
between both determinations point out the small effects of the new
data selection and reinforce the results obtained
in~\cite{GarciaMartin:2011jx}.
\begin{table}[h]
  \centering
\begin{footnotesize}
  \begin{tabular}{ccc}\hline
    &old UFD(MeV)&new UFD (MeV)\\\hline
    $\sqrt{s_{f_0(500)}}$&$(482\pm 16)-i(268\pm 16)$&$(483\pm 17)-i(266\pm 16)$\\\\
    $\sqrt{f_0(980)}$&$(1001\pm 3)-i(8\pm 9)$&$(1001\pm 5)-i(12\pm 10)$\\\\
    $\sqrt{\rho(770)}$&$(764.4\pm 2.0)-i(73.3\pm 1.2)$&$(764.2\pm 1.9)-i(73.2\pm 1.2)$
  \end{tabular}
\end{footnotesize}
  \caption{$f_0(500)$, $f_0(980)$ and $\rho(770)$ resonance pole positions of the old and new UFD obtained 
    from the analytic continuation to the complex plane of the twice subtracted Roy equation~\cite{GarciaMartin:2011cn}.
    The tiny differences between both determinations strengthen the error analysis carried out in~\cite{GarciaMartin:2011cn}.}
  \label{tab:poles}
\end{table}

\section{Elements of a self-consistent fit}
\label{sec:self-consistent}

As we have mentioned and demonstrated in the previous section, the
selection of data is an essential ingredient for a self-consistent
fit, namely a fit where residuals follow a normal
distribution. However, this selection has been done by hand, and a
more general procedure would be most useful. Guided by previous
experience in NN scattering~\cite{Perez:2014yla,Perez:2014kpa} we try
to explore a similar selection method.

Of course, one important aspect is that, unless specifically proven
incorrect, {\it all} available data as they are published by the
experimentalists should be taken into account to make the data selection.
However let us remember that, in the particular case of $\pi\pi$ scattering,
phase shift and inelasticity data are not obtained from direct measurements, 
but indirect, requiring the use of models, and thus, involve systematic uncertainties.  
Note by instance, that the PY05 average~\cite{Pelaez:2004vs} used in both,
the conformal and intermediate-energy parametrizations of~\cite{GarciaMartin:2011cn}
was obtained confronting these experimental analyses against Forward dispersion relations.  

Nevertheless, if in general we assume from the start that all experiments are
correct; our goal is to reach a consensus among {\it published}
experimental data which are or could be mutually compatible.  This is
an important assumption, but it is the most objective one we can make
without any detailed knowledge on how the experimental analysis was
carried out; if we {\it did} know, we would decide based on our own
judgement. We hope, however, that given a sufficiently large number of
data, statistical analysis will help us to make a judicious choice and
to reach a consensus among individual data measurements. In this way, we
envisage the possibility that simple experimental values in a given
experiment can individually be discarded without questioning the whole
experiment. A practical way to implement this method is to follow 
the $3\sigma$ self-consistent procedure proposed in~\cite{Perez:2013jpa},
which is given by the following three-steps algorithm: 
i) Make a first and global fit, ii) Discard from all data
those fulfilling $R_i^2 > 9$ iii) Re-fit the remaining data and
go to step ii) until convergence.  This selection method will generate
a boundary between accepted and rejected data and there will be a flow
of data across the boundary during the iterative process.

The good feature of this method is that in the case of two mutually
incompatible data it helps to decide which one is better suited even
when initially and individually one would reject both. Of course, this
decision is controlled by the fitting theory and unforeseen
restrictions on the theory may induce an undesirable bias in our
choice. It is thus important that the fitting curve is flexible enough
to prevent this situation~\footnote{From this point of view it is
  preferable to have an excess of fitting parameters, since their
  redundancy will emerge through correlations among them.  In the
  opposite situation, a too restrictive choice may reject data just on
  a lack of flexibility. Note that we want to use the most general and
  admissible theory which can accommodate all possible but correct
  data.}. Note that after this process has been carried out, we have
still to check whether the residuals pass the normality test.

In order to check the disagreement between the $\pi\pi$ data sets 
considered in the previous sections,
we apply the self-consistent approach to the S0 wave phase shift, 
performing a global fit to the data sets described in Sect.\ref{subsec:low} and \ref{subsec:inter}.
The advantage of a global fit, instead of considering separately the low and energy region, 
is that we can treat on an equal footing both energy regions, 
and thus, without introducing a priory any bias into our analysis. 
Note, however, that the hierarchy imposed in the original work of~\cite{GarciaMartin:2011cn},
allows to take advantage of the high precision accomplished in the latest NA48/2~\cite{ULTIMONA48}. 
 
In all, we find that if we fit the data in the S0 up to $\sqrt{s}=1.42
\, {\rm GeV}$ convergence is achieved already with two iterations,
with a total of only three points being discarded, 
namely, the ``old'' K decay data point~\cite{Rosselet:1976pu} at $\sqrt s=381.4$ MeV, (Table~\ref{tab:Conformal-residuals})
and the Kaminski et al.~\cite{Kaminski} data points at $\sqrt s=\{970, 1050\}$ MeV, (Table~\ref{tab:polynomial-residuals}). 
In Table~\ref{tab:S0parameters:self}, we provide the
resulting parameters in the first and second iterations. 
Note that the differences between the new UFD parameters and the original ones Table~\ref{tab:S0parameters}
are due to the global fit we are performing in the self-consistent method. 

As compared to the more detailed discussion in the previous section,
it is fair say that the $3\sigma$ self-consistent approach is much
simpler and effective, and does not require from the theoretician's
side a detailed discussion of the published data but rests on the
assumption that {\it most} of data are correct~\footnote{However, 
we stress again that in the case of $\pi\pi$-scattering,
where systematics dominate the experiment analyses,
dispersion relations allow to discard data sets 
which are in contradiction with the dispersive constraints~\cite{Pelaez:2004vs}.} 
 and that statistical regularity does imply a consensus among the mutually compatible
data. This makes sense of course provided the theoretical model used
to undertake the analysis is flexible enough.

Since the selection of data implies a sharp cut for residuals with
$R_i^2>9$, a comparison with the standard normal distribution is not
appropriate; the empirical distribution will have no tails. However it
is possible to compare against a \emph{truncated} normal distribution
$ \hat{N}(0,1)$ with $\hat{N}(x) = 0$ for $|x| > 3$. A first
comparison between this truncated distribution and the resulting
residuals has been made by looking at the corresponding moments. Of
course the values in Table~\ref{tab:moments} are no longer valid due
to the subtraction of the tails. The moments $\mu_r$ and $\Delta \mu_r$
for a finite 
size sample of the $3\sigma$-truncated distribution are shown in Table~\ref{tab:moments_trunc}. The results for $N=88$ along with the empirical
values for the residuals are shown in table~\ref{tab:3sigma_moments}. 

\begin{table}[h]
\begin{ruledtabular}
\begin{scriptsize}
\begin{tabular}{l|ccccccc}
 $r$ & 0 & 1 & 2 & 3 & 4 & 5 & 6   \\
 $\mu_r$ & 1. & 0. & 0.973 & 0. & 2.680 & 0. & 11.240 \\
 $\Delta \mu_r$ &
$ 0$ & $\frac{0.986}{\sqrt{\text{N}}}$ &
   $\frac{1.316}{\sqrt{\text{N}}}$ & $\frac{3.352}{\sqrt{\text{N}}}$ &
   $\frac{7.215}{\sqrt{\text{N}}}$ & $\frac{18.928}{\sqrt{\text{N}}}$ &
   $\frac{47.331}{\sqrt{\text{N}}}$  
\end{tabular}
\end{scriptsize}
\end{ruledtabular}
\caption{The normalized moments $\mu_r$ of the standardized $3 \sigma$-truncated gaussian
  distribution with their mean standard deviation $\Delta \mu_r$ for a
  sample of size $N$. For normal distributed data $x_i \in \hat N(0,1)$
  we expect that to $1\sigma$ confidence level $\sum_{i=1}^N x_i^r/N=
  \mu_r \pm \Delta \mu_r$.}
\label{tab:moments_trunc}
\end{table}

\begin{table}[h]
  \centering
\begin{scriptsize}
  \begin{tabular}{cccccccc}
               & $\mu_0$  & $\mu_1$ &$\mu_2$ &$\mu_3$&$\mu_4$&$\mu_5$&$\mu_6$\\\hline
    $R_{\rm 3\sigma}$& 1&-0.11& 0.59&-0.58&1.75&-3.41&9.36\\
    $\hat{N}(0,1)$ & $1\pm0$&$0\pm0.11$&$0.97\pm0.14$&$0.0\pm0.36$&$2.7\pm0.8$&$0\pm2.1$&$11.24\pm 5$\\
  \end{tabular}
\end{scriptsize}
  \caption{Central moments of the distribution of residuals obtained
    for the global self-consistent fit versus the moments of $N=88$
    random points normally distributed with truncation at $|x|=3$.
    $\mu_0$ and $\mu_1$ are by definition 1 and 0
    respectively. However from $\mu_2$ to $\mu_5$, deviations appear
    for empirical distribution. The deviations are larger for odd
    moments given the large asymmetries of the residuals.  For $n=6$,
    the values are compatible due to higher uncertainties.}
  \label{tab:3sigma_moments}
\end{table}

It is notable the improbably low $\chi^2/\nu$ values for this new fit
shown in table~\ref{tab:S0parameters:self}. This may be a consequence
of only discarding data with a too large contribution to the $\chi^2$
(mostly underestimated errors). Data with overestimated error bars
remain in the fit and their too low contribution to the $\chi^2$ pulls
down the final value. This actually shows that a selection of data
does not necessarily complies with (truncated) normality.

\begin{table}[h]
  \centering
  \begin{scriptsize}
  \begin{tabular}{ccc}
S0 wave   & Fit1 & Fit2 \\
\hline
$ B_0$ &$6.32\pm0.28$ &$6.10\pm0.27$ \\
   $B_1$& $-21.6\pm0.6$&$-20.1\pm0.6$\\
 $B_{2}$&$-32.4\pm1.5$&$-31.9\pm1.4$\\
 $B_{3}$&$-43.9\pm3.3$&$-50.8\pm3.2$\\
 $z_0$ & $M_\pi$& $M_\pi$\\
\hline
$d_0$ & $(227.2\pm1.3)^\circ$& $(228.8\pm1.3)^\circ$\\
$c$ & $(-675\pm  248)^\circ$& $(-195\pm  248)^\circ$\\
$B$ &  $(93.7\pm 2.3)^\circ$& $(92.2\pm 2.3)^\circ$ \\
$C$ & $(49.4\pm 2.9)^\circ$& $(52.2\pm 2.9)^\circ$ \\
$D$ & $(-86.4\pm 4.0)^\circ$& $(-89.8\pm 4.0)^\circ$ \\
\hline
N & 88 &  85\\
$\chi^2/\nu$ & 0.54 & 0.38\\
\hline
  \end{tabular}
  \end{scriptsize}
  \caption{S0 wave parameters for the self consistent global fit.
    Note that in this case, the whole phase shift energy region has been fitted at the same time.
    Fit 1 and Fit 2 stands for first and second iterations, indicating also the
    number of fitted data $N$ and the corresponding $\chi^2/\nu$
    (expected value is $1\pm \sqrt{2/\nu}$. Convergence is achieved in
    the second iteration.The first four lines correspond to the low
    energy parametrization, $\sqrt{s}\leq 0.85\,\mathrm{GeV}$, and the
    last nine to the parametrization up to
    $\sqrt{s}=1.42\,\mathrm{GeV}$.}
  \label{tab:S0parameters:self}
\end{table}

\section{Conclusions}
\label{sec:concl}

One of the most relevant accomplishments in hadronic physics in the
last decade has been the reliable determination of the mass and width
of the lowest scalar $0^{++}$ resonance, also known as the
$\sigma$-meson, which has entered into the PDG as the $f_0(500)$
state. This has occurred as a side-product of comprehensive long-term
studies in $\pi\pi$ scattering, a particularly simple reaction where
many theoretical constraints such as crossing, analyticity, unitarity,
chiral symmetry and Regge behavior allow for convincing and accurate
theoretical predictions. This conclusion holds after a long tour the
force, and thus the significance of this major issue should not be
underestimated.  In the present work we have re-analyzed the
statistical treatment of $\pi\pi$ scattering data from the point of
view of normality tests; which have so far been overlooked.
The basic aim was to check whether the currently accepted and agreed
bench-marking analyses carried out during the last decade fulfill the
condition that the difference between the fitted data and the fitting
theoretical curve can be regarded as a statistical fluctuation.  This
is an elementary requirement on the applicability of statistical
methods such as the least squares fit which can only be carried out
{\it a posteriori}.   
As an example, we have applied several conventional normality
tests to the  S0 wave of the precise $\pi\pi$ analysis performed in~\cite{GarciaMartin:2011cn}.
This test has pointed out a {\it tiny} violation of the normality requirements.
When the normality test fails many questions should be asked, but the most
obvious ones address either the compatibility of the data base used to
carry out the $\chi^2$-fit or the incapability of the theory to
describe the data or both. 
 We have carried out a preliminary analysis along these lines. While our study
can definitely be improved, we have analyzed several strategies
incorporating many of the elements that a full analysis might contain,
including data selection and normality tests.
However, the small differences obtained suggests that there is no need for 
a new reanalysis of $\pi\pi$ scattering.

We find that by changing slightly the selection of data, normality of
residuals can be achieved in a significant way. This allows to
propagate errors and hence to reassess the estimation of uncertainties
in the scalar resonance parameters. We find small and compatible
changes, reinforcing ultimately the benchmarking determinations
carried out during the last decade, and in particular the results
performed in~\cite{GarciaMartin:2011cn}.

\section*{Acknowledgments}

We would like to thank useful and productive discussions with J. R. Pelaez.
This work is supported by Spanish DGI (grant FIS2011-24149)
and Junta de Andaluc{\'{\i}}a (grant FQM225) 
and the DFG (SFB/TR 16, ``Subnuclear Structure of Matter'').
R.N.P. is supported by a
Mexican CONACYT grant.

\appendix 

\section{Data and residuals}

\begin{table}[h]
  \centering
  \begin{scriptsize}
  \begin{tabular}{lcccc}
Experiment   & $\sqrt{s}$ &$\left(O_i^{\rm exp}\pm\Delta O_i^{\rm exp}\right)^{(\circ)}$ &$O_i^{\rm th\,(\circ)}$&$R_i$\\
\hline
NA48/2 & 286.1 & $2.32\pm 1.89$ & 2.91&-0.31\\
       & 295.1 & $4.67\pm 1.26$ & 4.83&-0.12\\
       & 304.9 & $6.20\pm 0.95$ & 6.30&-0.11\\
       & 313.5 & $7.57\pm 0.82$ & 7.65&-0.10\\
       & 322.0 & $8.70\pm 0.64$ & 8.96&-0.41\\
       & 330.8 & $9.99\pm 0.58$ & 8.96&-0.55\\
       & 340.2 & $12.44\pm0.75$ & 11.75&0.91\\
       & 350.9 & $13.69\pm 0.57$ & 13.44&0.44\\
       & 364.6 & $16.60\pm 0.55$ & 15.62&1.78\\
       & 389.9 & $19.43\pm 0.50$ & 19.83&-0.80\\\hline
K-decays & 285.2&$-1.63\pm2.28$&2.73&-1.91\\
         & 289.0&$3.36\pm7.44$&3.44&-0.02\\
         & 299.5&$3.36\pm1.42$&5.43&-1.46\\
         & 303.0&$11.53\pm4.00$&6.00&1.38\\
         & 311.2&$7.25\pm1.08$&7.31&-0.06\\
         & 317.0&$7.08\pm2.85$&8.19&-0.39\\
         & 324.0&$8.87\pm0.96$&9.27&-0.41\\
         & 335.0&$11.29\pm2.28$&10.95&0.15\\
         & 340.4&$12.05\pm0.84$&11.80&0.29\\
         & 367.0&$15.80\pm2.77$&16.02&-0.09\\
         & 381.4&$16.86\pm1.18$&18.40&-1.31\\\hline
PY05 & 810&$88\pm6$&86.36&0.27\\
     & 830&$92\pm7$&89.20&0.27\\
     & 850&$94\pm6$&92.19&0.27\\
     & 870&$91\pm9$&95.5&-0.50\\
     & 910&$99\pm6$&104.0&-0.83\\
     & 912&$103\pm8$&104.6&-0.19\\
     & 929&$112.5\pm13.0$&110.1&0.18\\
     & 935&$109\pm8$&112.6&-0.45\\
     & 952&$126\pm16$&121.8&0.26\\\hline
  \end{tabular}
\end{scriptsize}
  \caption{Experimental S0 phase shift data used for the low-energy
    conformal fit, together with the fitted values and the residuals
    obtained from Eq.~\ref{residual-definition}. The experimental data
    can be gathered in three different blocks. The newest $K_{\ell4}$
    data from NA48/2~\cite{ULTIMONA48}, which incorporates a rigorous
    systematic and statistical error analysis, $K_{\ell4}$ data from
    old experiments~\cite{Rosselet:1976pu}, and the average result
    collected in PY05~\cite{Pelaez:2004vs}.}
  \label{tab:Conformal-residuals}
\end{table}

\newpage 

\begin{table}
  \centering
  \begin{scriptsize}
  \begin{tabular}{lcccc}
Experiment   & $\sqrt{s}$ &$\left(O_i^{\rm exp}\pm\Delta O_i^{\rm exp}\right)^{(\circ)}$ &$O_i^{\rm th\,(\circ)}$&$R_i$\\
\hline
PY05 & 870&$91\pm9$&95.5&-0.50\\
     & 910&$99\pm6$&104.0&-0.83\\
     & 912&$103\pm8$&104.6&-0.19\\
     & 929&$112.5\pm13.0$&110.1&0.18\\
     & 935&$109\pm8$&112.6&-0.45\\
     & 952&$126\pm16$&121.8&0.26\\
     & 965&$134\pm17$&132.7&0.10\\
     & 970&$141\pm21$&138.4&0.14\\ \hline
Kaminski & 910&$101.0\pm9.1$&104.0&-0.22\\
         & 930&$113.2\pm7.2$&110.5&0.25\\
         & 950&$123.2\pm7.1$&120.5&0.19\\
         & 970&$110.1\pm10.4$&138.4&-1.81\\
         & 990&$205.2\pm7.5$&192.6&1.11\\
         & 1010&$233.5\pm21.3$&230.6&0.09\\
         & 1030&$178.5\pm47.9$&234.7&-0.78\\
         & 1050&$192.2\pm11.5$&239.1&-2.72\\
         & 1070&$257.0\pm11.9$&243.8&0.74\\
         & 1090&$246.7\pm20.3$&248.7&-0.06\\
         & 1110&$262.4\pm8.9$&251.4&0.81\\
         & 1130&$221.8\pm22.9$&253.8&-0.93\\
         & 1150&$267.4\pm16.1$&256.3&0.46\\
         & 1170&$274.6\pm18.2$&259.1&0.57\\
         & 1190&$258.0\pm21.9$&262.2&-0.13\\
         & 1210&$265.2\pm13.5$&265.5&-0.02\\
         & 1230&$255.7\pm11.5$&269.2&-0.78\\
         & 1250&$258.2\pm14.8$&273.1&-0.67\\
         & 1270&$258.5\pm11.7$&277.4&-1.08\\
         & 1290&$273.2\pm10.9$&282.0&-0.53\\
         & 1310&$297.3\pm9.3$&286.9&0.74\\
         & 1330&$279.6\pm16.1$&292.2&-0.52\\
         & 1350&$305.6\pm10.3$&297.9&0.51\\
         & 1370&$296.9\pm21.7$&303.9&-0.32\\
         & 1390&$321.8\pm17.8$&310.4&0.13\\
         & 1410&$321.2\pm14.2$&317.3&-0.36\\\hline
Hyams & 1020&$230.2\pm6.4$&232.6&-0.38\\
      & 1060&$242.1\pm6.4$&241.4&0.10\\
      & 1100&$246.6\pm6.2$&250.5&-0.63\\
      & 1140&$255.3\pm6.1$&255.0&0.04\\
      & 1180&$256.3\pm5.5$&260.6&-0.78\\
      & 1220&$264.9\pm5.3$&267.3&-0.45\\
      & 1260&$274.1\pm5.8$&275.2&-0.19\\
      & 1300&$283.9\pm5.5$&284.4&-0.09\\
      & 1340&$289.1\pm5.3$&295.0&-1.11\\
      & 1380&$309.6\pm5.2$&307.1&0.47\\
      & 1420&$322.2\pm6.7$&320.9&0.17\\\hline
Grayer & 991.7&$231.3\pm13.5$&212.8&1.35\\
       & 1002.3&$206.6\pm20.1$&229.1&-1.11\\
       & 1031.7&$225.0\pm22.5$&235.1&-0.45\\
       & 1051.0&$245.3\pm12.1$&239.4&0.49\\
       & 1070.8&$249.8\pm10.3$&244.0&0.56\\
       & 1091.7&$251.6\pm7.7$&249.1&0.31\\
       & 1112.5&$251.6\pm10.3$&251.8&-0.03\\
       & 1133.3&$257.9\pm12.1$&254.2&0.30\\
       & 1150.0&$259.6\pm12.1$&256.3&0.27\\
       & 1173.3&$261.9\pm10.3$&259.6&0.22\\
       & 1193.3&$259.6\pm8.1$&262.7&-0.38\\
       & 1208.0&$270.0\pm6.0$&265.2&0.81\\
       & 1225.6&$271.0\pm7.4$&268.3&0.36\\
       & 1246.4&$282.3\pm6.5$&272.4&1.53\\
       & 1264.4&$278.2\pm8.6$&276.2&0.23\\
       & 1290.4&$285.1\pm8.2$&282.1&0.38\\\hline
  \end{tabular}
\end{scriptsize}
  \caption{Experimental S0 phase shift data used for the
    intermediate-energy polynomial fit, together with the fitted values
    and the residuals obtained from Eq.~\ref{residual-definition}. The
    experimental data sets considered are the PY05
    average~\cite{Pelaez:2004vs}, the re-analysis performed by
    Kaminski et al.~\cite{Kaminski}, the CERN-Munich data of the
    Solution. B of Grayer et al.~\cite{Grayer}, and the results given
    by Hyams et al.~\cite{Hyams75}, to which $5^{(\circ)}$ were added
    as systematic uncertainty.}
  \label{tab:polynomial-residuals}
\end{table}

\begin{table}
  \centering
  \begin{scriptsize}
  \begin{tabular}{lcccc}
Experiment   & $\sqrt{s}$ &$\left(O_i^{\rm exp}\pm\Delta O_i^{\rm exp}\right)^{(\circ)}$ &$O_i^{\rm th\,(\circ)}$&$R_i$\\
\hline
Protopopescu & 1000&$0.37\pm0.08$&0.41&-0.54\\
             & 1040&$0.38\pm0.04$&0.38&-0.07\\
             & 1075&$0.45\pm0.04$&0.48&-0.67\\
             & 1105&$0.50\pm0.04$&0.55&-1.36\\
             & 1135&$0.55\pm0.04$&0.62&-1.91\\
             & 1150&$.0.60\pm0.04$&0.66&1.54\\\hline
Hyams & 969.2&$0.68\pm0.38$&1.00&-0.83\\
      & 994.0&$0.08\pm0.08$&0.57&-6.24\\
      & 1008.8&$0.08\pm0.14$&0.36&0.15\\
      & 1028.6&$0.38\pm0.22$&0.36&0.39\\
      & 1048.4&$0.61\pm0.38$&0.40&0.53\\
      & 1068.1&$0.84\pm0.18$&0.46&2.14\\
      & 1087.9&$0.82\pm0.21$&0.52&1.43\\
      & 1107.7&$0.79\pm0.22$&0.56&1.04\\
      & 1127.5&$0.97\pm0.21$&0.61&1.74\\
      & 1147.3&$0.88\pm0.20$&0.66&1.13\\
      & 1167.0&$0.87\pm0.22$&0.70&0.77\\
      & 1186.8&$0.88\pm0.20$&0.74&0.72\\
      & 1206.6&$0.92\pm0.21$&0.77&0.71\\
      & 1226.4&$1.00\pm0.21$&0.80&0.94\\
      & 1246.2&$0.79\pm0.15$&0.83&-0.25\\
      & 1265.9&$0.84\pm0.14$&0.85&-0.04\\
      & 1285.7&$0.95\pm0.16$&0.86&-0.52\\
      & 1305.5&$0.82\pm0.16$&0.88&-0.37\\
      & 1325.3&$0.87\pm0.16$&0.88&-0.10\\
      & 1364.8&$0.82\pm0.16$&0.89&-0.47\\
      & 1384.6&$0.95\pm0.17$&0.89&0.34\\
      & 1404.4&$0.91\pm0.25$&0.89&0.08\\\hline
  \end{tabular}
  \end{scriptsize}
  \caption{Experimental S0 inelasticity data, together with the fitted
    values and the residuals obtained from
    Eq.~\ref{residual-definition}. Only data sets from Hyams et
    al. and Protopopescu et al.~\cite{pipidata}, which are consistent
    with the dip solution, are considered. With the exception of the
    data of Kaminski et al.~\cite{Kaminski} as it is explained in the
    text above.}
  \label{tab:eta-residuals}
\end{table}

\end{document}